%% file: ms.tex
\documentclass[runningheads]{llncs}

\usepackage{amsmath}
\usepackage{amssymb}
\usepackage{float}
\usepackage{subcaption}
\usepackage{petriGames}

\begin{document}
\title{High-Level Representation of \\
    Benchmark Families for Petri Games
\thanks{This work was supported by the German Research Foundation (DFG) through the grant Petri Games (No.\ 392735815)}}

\titlerunning{High-Level Representation of Petri Games}

\author{Manuel Gieseking \and Ernst-R\"udiger Olderog}

\authorrunning{M.\ Gieseking and E.-R.\ Olderog}

\institute{University of Oldenburg
\email{\{gieseking, olderog\}@informatik.uni-oldenburg.de}}

\maketitle

\begin{abstract}
Petri games have been introduced as a multi-player game model representing
causal memory to address the synthesis of distributed systems. For Petri 
games with one environment player and an arbitrary bounded number of 
system players, deciding the existence of a safety strategy is 
EXPTIME-complete.
This result forms the basis of the tool \adam{} that 
implements an algorithm for the synthesis of distributed controllers from 
Petri games.
To evaluate the tool, it has been checked on a series of 
parameterized benchmarks from manufacturing and workflow scenarios. 

In this paper, we introduce a new possibility to represent benchmark families 
for the distributed synthesis problem modeled with Petri games. It enables 
the user to specify an entire benchmark family as one parameterized high-level net.  
We describe example benchmark families
as a high-level version of a Petri game and exhibit an instantiation yielding 
a concrete 1-bounded Petri game.
We identify improvements either regarding the size or the functionality
of the benchmark families by examining the high-level Petri games.
\end{abstract}

\section{Introduction}
\label{sec:introduction}
Automatically creating a program from a formal specification without any human programming involved,
 is of great interest for the implementation of correct systems.
A \emph{synthesis} algorithm either automatically derives an implementation 
satisfying a given formal specification
or states the non-existence of such an implementation \cite{Church/57/Applications}.
For \emph{reactive systems}, i.e., system which continuously interact with their environment, the synthesis problem is 
often described as a game between the environment and the system.
In this game-theoretic approach the specification is given as a \emph{winning condition} of the game
and a correct implementation is a \emph{strategy} for the system players
which satisfies the given winning condition against all moves of the environment.
The synthesis approach fundamentally simplifies the development of complex systems
by defining only the possible actions of the system and specifying the winning condition over these actions.
This puts the development process on a more abstract level and avoids the error-prone manual coding.

For the \emph{monolithic synthesis}, where the system can be seen as one unit with a central controller as strategy,
there is a growing number of tools~\cite{Jobstm07c,DBLP:conf/tacas/Ehlers11,BGHKK12,DBLP:conf/cav/BohyBFJR12} solving nontrivial applications.
However, for the synthesis of \emph{distributed systems}, i.e.,
systems composed of multiple independent processes possibly distributed over wide distances,
the tool support is restricted.
This is mainly due to the high complexity of the solving algorithms
or the undecidability results for the general problem.
In the two well-established models, 
the \emph{Pnueli/Rosner} model~\cite{Pnueli+Rosner/90/Distributed} and \emph{Zielonka's asynchronous automata}~\cite{DBLP:journals/ita/Zielonka87}, the complexity is in general
nonelementary~\cite{gasLerZei05,walu13,DBLP:conf/fsttcs/MadhusudanTY05} or even undecidable~\cite{Pnueli+Rosner/90/Distributed,FinkbeinerSchewe05}.
For the class of Zielonka automata with acyclic communication architectures
the control problem has been shown to be decidable,
with non-elementary complexity in general but EXPTIME for the special case
of architectures of depth 1~\cite{MuschollWalukiewicz2014}.
For \emph{Petri games}~\cite{FinkbeinerOlderog14,FinkbeinerOlderog16},
which have been shown to adequately model problems from manufacturing and 
workflow scenarios,
reasonable subclasses can be solved with affordable costs
and suitable tool support \cite{fingieold15,Finkbeiner15,fingiehecold17}.

In this paper, we extend the work on Petri games and present a model 
for representing benchmark families for the synthesis of distributed systems
in a concise way.
\emph{Petri games} model the distributed synthesis problem as a game between two teams:
the environment players, representing external influences (the \emph{uncontrollable} behavior),
and the system players, representing the processes (the \emph{controllable} behavior).
In Petri games each player is modeled as a token of an underlying place/transition Petri net.
The places of the net are partitioned between the teams.
All players remember their own \emph{causal} past and
communicate this knowledge to every player participating in a joint transition.
An example can be seen in \refFig{as}.

Benchmark families depend on parameters which define a set of problems with increasing complexity.
The new representation is based on schemata of Coloured Petri Nets~\cite{GenrichL81,jensen92}
rather than place/transition Petri nets, to resemble the parameters and sets of problems.
We use places with individual tokens ranging over predefined domains of parametric size,
transitions labelled with conditions that guard their firability,
and arcs labelled with expressions stating the result of the firing.
Conditions and expressions may have variables ranging over the predefined domains.
This enables the user to specify the entire benchmark family as one parametric high-level net rather
than introducing a set of instances of the family and 
descriptions how to generalize these Petri games.
Generally, the individual elements of a benchmark family (e.g., robots, work pieces, tools, humans, etc.)
can be modeled by parametric sets of individual tokens and are processed by the transitions
according to the semantics.
Figure~\ref{fig:asHL} serves as an example for a set of alarm systems and locations of a burglary.

In this paper,
we introduce a new parameterized high-level representation of Petri games 
based on high-level Petri nets for a concise and clear definition of benchmark families.
We apply the new definition to some of the existing benchmark families and 
show the correspondence of the high-level version to an example instantiation.
During the application we identified improvements (either in size or functionality)
of these benchmark families.

The remainder of the paper is structured as follows.
Section~\ref{sec:petriGamesRecap} recaps the ideas, results, and solving techniques of Petri games
and informally motivates the new high-level representation by an example.
The formal definition of the high-level representation is given in \refSection{parameterizedHighLevelPetriGames}.
In \refSection{benchmarkFamilies} we illustrate the new approach by presenting 
two examples from the manufacturing domain and depicting for each example
both the high-level representation and an instantiation.

\section{Petri Games for the Synthesis of Distributed Systems}
\label{sec:petriGamesRecap}
In this section a brief overview of 
Petri games~\cite{FinkbeinerOlderog14,FinkbeinerOlderog16} is given.
We illustrate the model via an instantiation of the benchmark family of an distributed alarm system from~\cite{fingiehecold17}
and motivate the new high-level representation for a concise and clear presentation of the family. 
Basic knowledge about \emph{Petri nets}~\cite{DBLP:journals/tcs/NielsenPW81,DBLP:books/sp/Reisig85a} is assumed.
We fix the notation of a Petri net \(\petriNet\),
with places \(\pl\),
transitions \(\tr\),
a flow relation \(\fl\subseteq(\pl\cup\tr)\times(\tr\cup\pl)\),
and an initial marking \(\init\subseteq\pl\).

\subsection{Petri Games}
\label{sec:petriGames}
A \emph{Petri game} \(\pGame=(\plS,\plE,\tr,\fl,\init,\bad)\) models the distributed synthesis problem as a multi-player game
where the tokens of an underlying Petri net \(\pNet\)
represent the players of the game. 
The players act in two teams: the uncontrollable players (\emph{environment players}) are the token residing on
environment places \(\plE\) (depicted as white circles) and controllable players (\emph{system players})
are the token residing on the system places \(\plS\) (depicted as gray circles).
Those sets are the disjunct union of the places of the underlying net, i.e., \(\pl=\plE\,\dot\cup\,\plS\).
The uncontrollable players are used for modeling external influences on the system,
whereas the controllable ones represent its processes.
Each player knows its own \emph{causal past}, i.e., the places and transitions which had been used to reach the current place.
This information is exchanged with all players participating at a joint transition.
These intricate causal dependencies (and independencies) are naturally represented by the
\emph{unfolding} of the Petri net~\cite{DBLP:series/eatcs/EsparzaH08,DBLP:journals/acta/Engelfriet91}.
An unfolding represents the behavior of a Petri net by unrolling each loop of the execution
and introducing copies of \(p\in\pl\) for each join of transitions in \(p\).
The system players have to cooperate to win the game, i.e.,
to avoid reaching certain \emph{bad places} \(p\in\bad\) (depicted as double circled places). 
To satisfy this \emph{safety objective}, the players can solely use their locally available information.

A \emph{strategy} is a local controller for each
system player which only decides on its current view and available information about the whole system.
A strategy can be obtained by removing certain branches of the unfolding.
That is, transitions and their complete future are removed which are considered as not be taken from the system.
We search for \emph{deterministic}, i.e., in every situation no two transition are enabled,
and \emph{deadlock-avoiding} strategies, i.e., whenever the system can proceed in \(\pGame\)
there must also be a continuation in the corresponding situation in the strategy.
Furthermore, no behavior of the environment is allowed to be restricted.

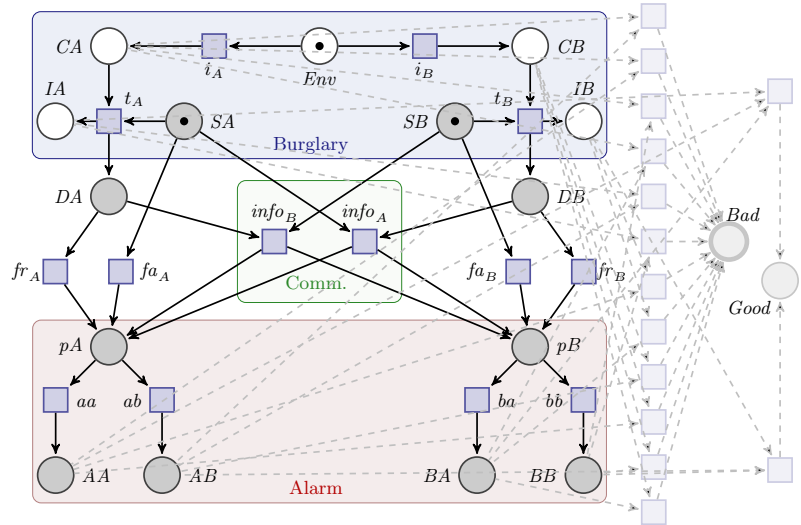
\begin{figure}[t]
	\centering
	\scalebox{0.8}{
		\input{as}
	}
	\caption{Two distant locations \(A\) and \(B\) are secured by the
	alarm systems represented by the token initially residing in \(\mathit{SA}\) and \(\mathit{SB}\).
	The alarm system in location \(X\) can state that there should be a burglary at location \(Y\) by 
	putting a token at place \(\mathit{XY}\) (for \(X,Y\in\{A,B\}\)).
	The goal is that no system produces a false alarm, or, in case of an intrusion, indicates the wrong intrusion point.}
	\label{fig:as}
\end{figure}
\begin{figure}[t]
	\centering
	\scalebox{0.7}{
		\input{asUnfolding}
	}
	\caption{Unfolding and winning strategy of the Petri game from \refFig{as}.
	The places \(\mathit{Bad}\) and \(\mathit{Good}\) and the corresponding transitions are omitted for readability reasons.
	The winning strategy for the system players is visualized by the solid elements.}
	\label{fig:asUnfolding}
\end{figure}
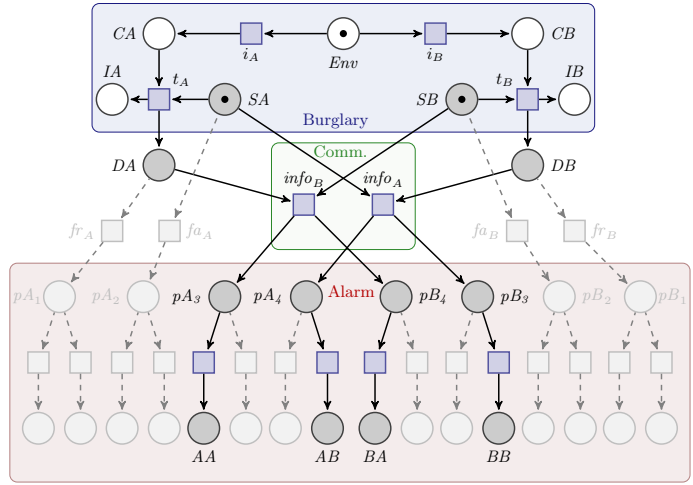

We illustrate the model with an example of two system players and one environment player,
modeling a distributed alarm system from \cite{FinkbeinerOlderog14}, visualized in \refFig{as}.
\begin{example}[Alarm System]
\label{ex:as}
We consider two alarm systems \(\mathit{SA}\) and \(\mathit{SB}\) securing one location each. 
Location \(A\) is depicted as the left and location \(B\) as the right part of \refFig{as}.
The alarm systems are represented by the tokens in the corresponding system places.
In case of a burglary at any of these locations,
executed by the environment token,
each alarm system should indicate the correct intruding point despite their distribution.
That is, for an intrusion in location \(Y\in\{A,B\}\) the token of each system \(X\in\{A,B\}\) should 
eventually reach \(\mathit{XY}\).
Each alarm system has in addition to its correct behavior 
the possibility to trigger a \emph{false alarm},
i.e., setting off an alarm without any burglary, 
or to give a \emph{false report}, i.e., indicating the wrong location of
the burglary.
These incorrect behavior can occur by taking transition \(\mathit{fa}_X\) or \(\mathit{fr}_X\) for \(X\in\{A,B\}\),
respectively.
Thus, each alarm system \(\mathit{SX}\) for \(X\in\{A,B\}\) should wait until a burglary has happened
and then inform the other system \(Y\) (via transition  \(\mathit{info}_Y\)),
or wait on getting informed by the other system (via transition \(\mathit{info}_X\)).
Generally, the system should only take a decision (and the right one)
when it is well enough informed.
This results in the strategy depicted as part of the unfolding in \refFig{asUnfolding} by the solid elements.
\end{example}

\subsection{Solving Petri Games}
\label{sec:decidingPetriGames}
There are four major results on finding winning strategies for Petri games with safety objectives.
Firstly, deciding the question whether it exists a strategy for the system players for Petri games with
one environment player and an arbitrary but bounded number of system players and a safety objective is EXPTIME-complete~\cite{FinkbeinerOlderog16}.
The strategy can be obtained in single-exponential time.
Secondly, interchanging the players, meaning the setting of \(n\in\N\) distributed environment players
and one system player, yields the same complexity results~\cite{DBLP:conf/fsttcs/FinkbeinerG17}.
Thirdly, for unbounded underlying Petri nets the question is undecidable~\cite{FinkbeinerOlderog16}.
Finally, the paper~\cite{Finkbeiner15} introduces a bounded synthesis approach which limits the size of the strategy.
This constitutes a semi-decision procedure which is optimized in finding small implementations.

In the following we briefly recap the idea of the decision procedure
for one environment player and \(n\in\N\) system players on a Petri game with a safety objective
and an underlying 1-bounded Petri net.
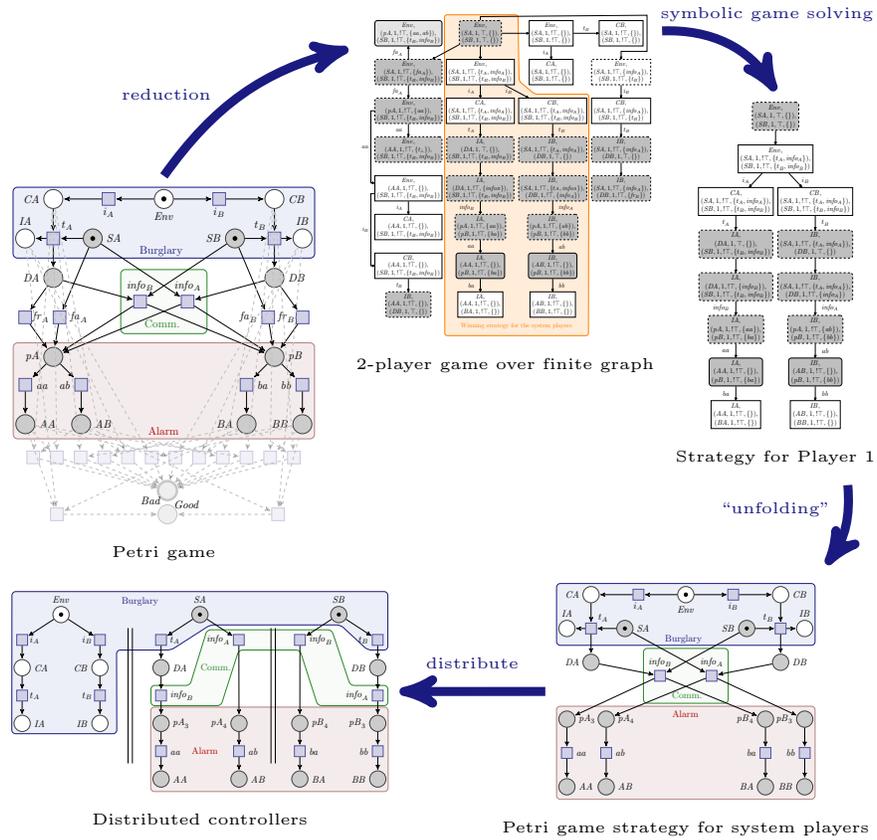
\begin{figure}[t]
	\centering
	\scalebox{1.2}{
		\input{solvingPetriGames}
	}
	\caption{An overview of the symbolic game solving algorithm for 1-bounded Petri games with one environment
	and an arbitrary number of system players with a safety objective implemented in \adam~\cite{fingieold15,fingiehecold17}.}
	\label{fig:solving}
\end{figure}
The algorithm consists of four major steps:
Firstly, the input Petri game is reduced to a two-player game over a finite graph \(\ifGame\) with complete information.
Secondly, the question of the existence of a strategy in \(\ifGame\) is answered with standard symbolic game solving algorithms
and a strategy for \(\ifGame\) is constructed.
Thirdly, the strategy of \(\ifGame\) is used to extract a common strategy for the system players of \(\pGame\).
Fourthly, this strategy is distributed into one local controllers for each process.
An overview of the general approach is visualized in Fig.~\ref{fig:solving}. 

The algorithm starts with a Petri game in the upper left corner,
which consists of one environment, a bounded number of system players,
and places denoted as bad.
This net is reduced to a two-player game over a finite graph with complete information, 
i.e., both players know in any point in time everything about the opponent.
Player 0 (depicted as the white rectangles) represents the one environment player and
Player 1 (depicted as the gray rectangles) represents all system players together.
The key idea of the reduction is that the behavior of the environment player is delayed 
until no system player can move without any interaction with the environment (or never depend on the environment anymore).
This ensures that we can consider the players as completely informed about all actions 
in the game. The system players
will be informed of the environment action by their next movement and they
are also informed of the other system player's behavior because deterministic strategies are build.
A two-player game over a finite graph with complete information can be solved 
with standard game solving techniques.
Furthermore, the existence
of a strategy already yields the existence of a \emph{memoryless} strategy,
i.e., a strategy which is only dependent on the current state and not on the previous states of the run.
In \cite{fingieold15} this is done with a symbolic game solving algorithm utilizing BDDs
for the representation of the state space.
In \cite{FinkbeinerOlderog14} it is shown that a strategy for the system players of the 
Petri game exists
if and only if a strategy for Player 1 exists in the two-player game.
Thus, we achieve a memoryless strategy for the system players such that they can cooperatively play
without encountering any bad behavior against all possible actions of an hostile environment.
By traversing the winning strategy of the two-player game over the finite graph in breadth-first order,
a finite Petri net can be constructed which is a winning strategy of the system players of the Petri game.
Finkbeiner and Olderog \cite{FinkbeinerOlderog14} showed 
that for a \emph{concurrency-preserving} strategy,
i.e., the number of ingoing arcs is equal to the number of outgoing arcs for each transition,
this common strategy of the system players can be distributed into local strategies.
This yields one controller for each player.

\subsection{Motivating the High-Level Representation}
As for the comparison of standard P/T Petri nets and high-level Petri nets,
the beauty of the high-level representation explicated in the following section
consists of the conciseness and clarity of the illustration of the system's behavior.
This can be seen in \refFig{asHL}.
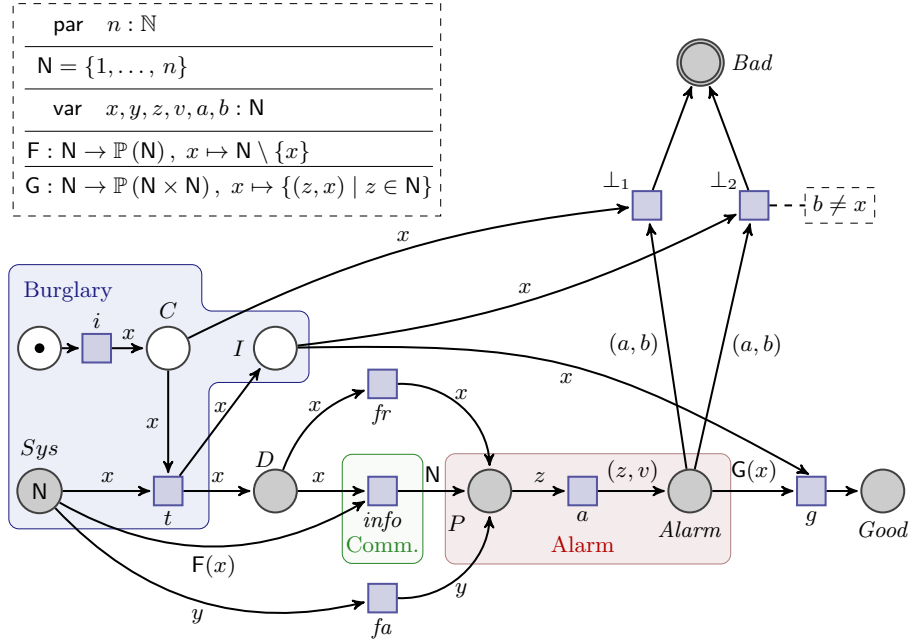
\begin{figure}[t]
	\scalebox{0.95}{
		\input{asHL}
	}
	\caption{Parameterized high-level Petri game for a benchmark family with \(n\in\N\) distributed alarm systems.
	The low-level Petri game of Fig.~\ref{fig:as} can be seen as an instantiation with \(n=2\)
	of the benchmark family.}
	\label{fig:asHL}
\end{figure}
In \cite{fingiehecold17} the Petri game of the two alarm systems of Example~\ref{ex:as} is extended to a benchmark family
of \(n\) alarm systems which all have to be informed about the burglary and set off the alarm accordingly.
This means that new intrusion points and new alarm systems are introduced by adding copies of the 
corresponding places and transitions.
Using P/T Petri nets only, the exact description of a benchmark family requires 
a high amount of precise descriptive texts
but nevertheless bears the risk of introducing misunderstandings.
Also an instantiation, e.g., the one in \refFig{as},
cannot show the details appropriately.
For example, it does not imply whether in this
benchmark family the burglarized alarm system informs the others synchronously or in any manner sequentially.
But visualizing the family for three systems is already quite unwieldy; especially for the transitions leading into the bad place.
The representation in \refFig{asHL}, which syntax and semantics definitions are given in the following section,
however allows for a concise, parametric definition of the benchmark family.

We introduce the concept by the example presented in \refFig{asHL} before providing the technical details in the following section.

\begin{example}[Parameterized Alarm System]
\label{exam:asHL}
In the high-level representation individual tokens can reside on places and the transitions 
can move them individually. Furthermore, the firing of transitions can be restricted by guards.
In the example, \(n\) individual tokens, with \(n\in\N\), 
representing \(n\)  alarm systems initially reside in the place \(\mathit{Sys}\).
The burglar,
represented by the black token, can intrude any of the \(n\) locations
(represented by putting a token \(x\in{\sf N}\) into the place \(C\))
via transition \(i\).
Only the corresponding alarm system at that location
can detect the intrusion by transition \(t\), because both ingoing arcs are labelled 
with the same identifier \(x\).
In any case, every alarm system can set off a false alarm by transition~\(\mathit{fa}\).
After detecting an intrusion, an alarm system can synchronously inform all other systems by transition \(\mathit{info}\)
or do not report the intrusion by \(\mathit{fr}\).
Finally, it can decide by transition \(a\) which alarm to set off.
It is \emph{bad} if one alarm system \(z\) decides to set off an alarm for location \(v\), 
by putting \((z,v)\) into \(\mathit{Alarm}\)
but another location has been intruded (transition \(\bot_2\) leading to 
place \(\mathit{Bad}\)) 
or if some alarm is set off, but no intrusion has ever been detected 
(transition \(\bot_1\) leading to place \(\mathit{Bad}\)).
If all alarm systems have detected the intruded location correctly, 
the place \(\mathit{Good}\) can be reached.
\end{example}

By replacing the transition \(\mathit{info}\) according to \refFig{sequentiallyComm},
we can easily switch from a synchronously informing of the other systems 
to an arbitrary sequential order of information dissemination.
\begin{figure}[t]
\centering
	\begin{subfigure}[t]{0.48\textwidth}
		\centering
		\begin{tikzpicture}[node distance=2cm and 1.5cm,>=stealth',bend angle=45,auto,on grid]
			\node [sysplace]			 		(sx)        	[label=above:\(Sys\)]	                         	 {\({\sf  N}\)};
			\node [sysplace, right =of sx,xshift=2mm]		(sxx)        	[label={[label distance=-1mm,xshift=-1.5mm]above:\(D\)}]                         	 {};
			\node [transition, right =of sxx,xshift=0mm]		(info)  	[label={[label distance=-0.5mm]below:$\mathit{info}$}]	 {};
			\node [sysplace, right =of info]	 		(px)        	[label=below:\(P\)]	                         	 {};

			\path[-latex, thick]
			 	(info)	edge[pre]				node[above] 		{\(x\)}                         (sxx)
					edge[pre, bend left=30]			node[below,pos=0.5] 		{\({\sf  F}(x)\)}                      (sx)
					edge[post]  				node[above]	 	{\({\sf  N}\)}            	(px)
			;
		\end{tikzpicture}
	\subcaption{Transition \(\mathit{info}\) of \refFig{asHL} for informing all other systems \emph{synchronously}.}
	\label{fig:comSynch}
	\end{subfigure}\hspace{3mm}
	\begin{subfigure}[t]{0.48\textwidth}	
		\centering
		\begin{tikzpicture}[node distance=2cm and 1.5cm,>=stealth',bend angle=45,auto,on grid]
			\node [sysplace]			 		(sxSeq)        	[label=above:\(Sys\)]	                         	 {\({\sf  N}\)};
			\node [sysplace, right =of sxSeq,xshift=2mm]		(sxxSeq)        	[label={[label distance=-1mm,xshift=-1.5mm]above:\(D\)}]                         	 {};
			\node [transition, right =of sxxSeq,xshift=0mm]		(infoSeq)  	[label={[label distance=-1.5mm]below right:$\mathit{info}$}]	 {};
			\node [sysplace, right =of infoSeq]	 		(pxSeq)        	[label=below:\(P\)]	                         	 {};
			
			\node [predicate, above =of infoSeq, yshift=-12mm]	(prA)									 {\(x\neq y\)};
			\draw [-,dashed, thick] (infoSeq) -- ([xshift=0.1mm]prA.south);

			\path[-latex, thick]
			 	(infoSeq)	edge[pre,bend right=20]			node[above] 		{\(x\)}                      (sxxSeq)
						edge[post, bend left=20]		node[below,pos=0.5] 	{\(y\)}                      (sxxSeq)
						edge[post]  				node[above]	 	{\(x\)}		              (pxSeq)
			;

			\path[-latex, thick]
			 	(infoSeq.south)	edge[pre, bend left=30]			node[below,pos=0.5] 	{\(y\)}                      (sxSeq)
			;
		\end{tikzpicture}
	\subcaption{A possible replacement for \(\mathit{info}\) to inform all other systems \emph{sequentially}.}
	\label{fig:comSeq}
	\end{subfigure}
\caption{The left figure shows the synchronously informing of the burglary of \refFig{asHL}.
The right one introduces a possible replacement of the transition \(\mathit{info}\),
such that after one system detected the intruding or got informed about the intrusion,
it informs an arbitrary other system. Thus, the systems can inform one another in any arbitrary order.
}
\label{fig:sequentiallyComm}
\end{figure}
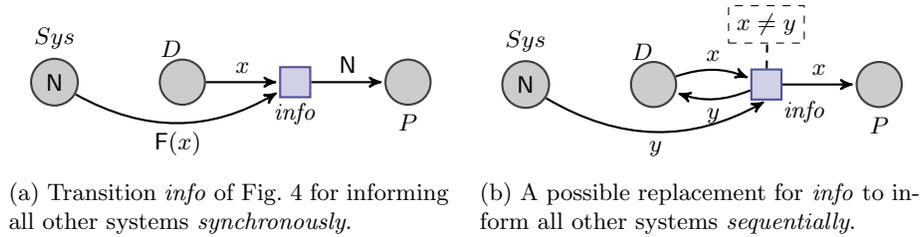
Note that the strategy of each alarm system can decide which other system should be informed next.
The last informed system can take transition \(\mathit{fr}\) because no other system has to be informed anymore.
This yields \(n\cdot(n-1)\) transitions in the low-level version, in contrast to the \(n\) ones 
in the synchronous case.
Such differences are more complex to visualize in the low-level presentation
because the difference can only be recognized for \(n>2\).

\section{Parameterized High-level Petri Games}
\label{sec:parameterizedHighLevelPetriGames}
In high-level Petri nets \emph{values} may appear as individual tokens 
in places \cite{Reisig2013}.
Such a value is also referred to as a ``color'', leading to the terminology of
Coloured Petri Nets \cite{jensen92}.
In high-level Petri nets, 
the ingoing and outgoing arcs of transitions are labelled by expressions
that specify which of the individual tokens are withdrawn from the preset
and which ones are added to the places in the postset of the transition.
Additionally, Boolean expressions labelling the transitions serve as guards.

In this section, we use these concepts to introduce high-level Petri games. 
We constrain ourselves to high-level Petri games that have sets
(rather than multisets) of individual tokens in their places.
We consider parameterized high-level games where the size of the sets
of individual tokens that may appear in the places depends on parameters.

\subsection{Preliminaries}
We consider \emph{parameters}, with typical letters $k,m,n$, ranging
over the set $\mathbb{N}$ of natural numbers and write
${\sf par}\ k,m,n : \mathbb{N}$ to declare that $k,m,n$ are parameters.
There may be a constraint added to the parameters like $m \le n$.
An \emph{instantiation} assigns
a fixed natural number to each parameter.
Parameters may appear in \emph{set expressions} $S$, defined inductively 
by the following syntax:
\[
 S ::= \{1,\dots, n\}\
                   |\ \{ \bullet \}\ 
                   |\ S_1 \times \dots \times S_n\
                   |\ \mathbb{P}(S)
\]
Here $\{1,\dots, n\}$ is a finite set of parametric size $n$,
the symbol $\bullet$ denotes the \emph{black token} used in normal Petri nets,
$\times$ denotes cartesian product, and $\mathbb{P}$ the power set.
Set expressions are used as (parametric) \emph{types}.
An instantiation of the parameters turns each set expression into
a fixed set.
\emph{Constants}, with typical letters {\sf K,M,N}, are used as
abbreviations for set expressions. We write ${\sf K} = S$ to declare
that {\sf K} abbreviates the set expression $S$.

We consider \emph{variables}, with typical letters $x,y,z$, ranging over set expressions
and write ${\sf var}\ x,y,z : S$ to declare that $x,y,z$ are variables of type $S$.
We write $\mathit{ty}(x)$ to denote the type of a variable $x$.
We consider \emph{function symbols}, with typical letters {\sf F,G},
and write ${\sf F}: S_1 \longrightarrow S_2$ to declare that
{\sf F} is a symbol standing for a function
from elements of $S_1$ to elements of $S_2$, for set expressions $S_1, S_2$.

Out of parameters, constants, variables, and function symbols
we construct Boolean expressions and expressions of set type.
We shall not define the syntax of these expressions in detail here,
but give typical examples. 
Suppose ${\sf par}\ m,n : \mathbb{N}$ and ${\sf var}\ x,y,x',y' : S_1$ and
${\sf F}: S_1 \longrightarrow S_2$.
Then  $m<n$, $x \neq y$, and $x = x' \land y = y'$ are Boolean expressions,
the pair $(x,y)$ is an expression of type $S_1 \times S_1$
and the function application ${\sf F}(x)$ is an expression of type $S_2$.
To define the function denoted by {\sf F} we write 
a maplet $x \mapsto e$, where $x$ is a variable of type $S_1$ and
$e$ is an expression of type $S_2$ containing $x$ as a free variable.
For a given instantiation, the maplet describes how {\sf F}
assigns to a given element $d$ of type $S_1$ a value of type $S_2$
by evaluating $e$ with $d$ substituted for $x$ in~$e$.

\subsection{High-level Petri Games}
In high-level Petri games \emph{values} may appear as individual tokens
in addition to the black tokens of normal Petri nets.
Syntactically, a high-level Petri game is a structure
\[
 \mathcal{H} = (\plS^H,\plE^H, \tr^H, \fl^H, \init^H, \bad^H,\mathit{ty},g,e, in),
\]
where the following components are as in 1-bounded Petri games:
\begin{itemize}
  \item $\plS^H$ is a set of \emph{system places},
        \vspace{1mm}
  \item $\plE^H$ is a set of \emph{environment places},
        \vspace{1mm}
  \item $\pl^H$ is the set of all places: $\pl^H = \plS^H\, \cup\, \plE^H$,
  \item $\tr^H$ is a set of \emph{transitions},
  \item $\fl^H \subseteq (\pl^H \times \tr^H) \cup (\tr^H \times \pl^H)$
        is the \emph{flow relation},
  \item $\init^H \subseteq \pl^H$ is the set of \emph{initially marked} places,
  \item $\bad^H \subseteq \pl^H$ is the set of \emph{bad places}. 
\end{itemize}
Additionally, the following components represent the high-level structure:
\begin{itemize}
  \item $\mathit{ty}$ is a mapping that assigns to each place $p \in \pl^H$
        a \emph{type} $\mathit{ty}(p)$ in the form of a set expression, 
        describing the set of individual tokens that may reside in $p$ 
        during the game,
        
  \item $g$ is a mapping that assigns to each transition $t \in \tr^H$
        a Boolean expression $g(t)$ serving as a \emph{guard}
        describing when $t$ can fire,
  
  \item $e$ is a mapping that assigns to each ingoing arc 
        $(p,t) \in \fl^H$ and each outgoing arc $(t,q) \in \fl^H$ 
        of a transition $t \in \tr^H$
        an expression $e(p,t)$ and $e(t,q)$ of set type, respectively,
        describing which tokens are withdrawn by $t$ from $p$ and 
        which tokens are placed by $t$ on $q$ when $t$ is fired,
        
  \item $\mathit{in}$ is a mapping that assigns to each initially marked place
        $p \in \init^H$ a non-empty subset of $\mathit{in}(p) \subseteq \mathit{ty}(p)$.
\end{itemize}
Guards and expressions will typically contain variables.
For a transition $t \in \tr^H$ let ${\sf var}(t)$ denote the set of
free variables occurring in the guard $g(t)$ or in one of the expressions
$e(p,t)$ and $e(t,q)$ for places $p$ in $t$'s  \emph{preset}, 
defined by $\mathit{pre}(t) = \{ p \in \pl^H \mid (p,t) \in \fl^H\}$, 
or $q$ in $t$'s \emph{postset}, 
defined by  $\mathit{post}(t)=\{ q \in \pl^H \mid (t,q) \in \fl^H\}$.

Graphically, a high-level Petri game $\mathcal{H}$ 
looks like a normal Petri game, except that guards $g(t)$
appear inside a dashed box connected to the transition $t$
by a dashed line, 
expressions $e(p,t)$ and $e(t,q)$ appear as labels of 
the arcs $(p,t)$ and $(t,q)$, respectively,
and types $\mathit{ty}(p)$ appear as labels of places $p$.
To avoid clutter, guards equivalent to \emph{true} are not shown.
Also, if the type of a place $p$ can be easily deduced from the context,
the label $\mathit{ty}(p)$ is not shown.
The declarations of parameters, constants, variables, and function 
symbols are listed in a dashed box near the graphics of the Petri game.

The \emph{semantics} of a high-level Petri game $\mathcal{H}$ is given by its token game.
To define it, we assume an instantiation of the parameters so that 
each set expression defines a fixed set.
A \emph{marking} $M$ of $\mathcal{H}$ assigns to each place $p$ a \emph{set} 
$M(p) \subseteq \mathit{ty}(p)$. Unlike in \cite{jensen92}, we do not admit multisets as markings
because we aim at 1-bounded Petri games as low-level instantiations of high-level 
Petri games.
The \emph{initial marking} $M_0$ of $\mathcal{H}$ is the marking with
$M_0(p) = \mathit{in}(p)$ for $p \in \init^H$ and
$M_0(p) = \emptyset$ otherwise.

A \emph{valuation} $v$ \emph{of a transition} $t$
assigns to each variable $x \in {\sf var}(t)$ a value $v(x) \in \mathit{ty}(x)$. 
By $\mathit{Val}(t)$ we denote the set of all valuations of $t$.
Each valuation $v$ of $t$
is lifted inductively from the variables in ${\sf var}(t)$ 
to the expressions around $t$.
For the guard $g(t)$ we denote by $v(t)$ 
the Boolean value assigned by $v$ to $g(t)$. 
For an ingoing arc $(p,t)$ we denote by $v(p,t)$ the value assigned 
by $v$ to $e(p,t)$, and analogously for an outgoing arc $(t,p)$.

A transition $t$ is \emph{enabled at a marking $M$ under a valuation $v$ of $t$} 
if $ v(t) = true$ and $v(p,t) \subseteq M(p)$ for each arc $(p,t)$.
\emph{Firing} (the enabled) transition $t$ at $M$ under $v$
yields the marking $M\,'$, where for each place $p$
\[
  M\,'(p) = (M(p) - v(p,t)) \cup v(t,p).
\]
This is denoted by $M\, \fire{t,v}\, M'$.
We assume here that $\cup$ is a disjoint union, which is satisfied 
if the Petri game is \emph{contact-free}, i.e., if for all $t \in \tr^H$
and all reachable markings $M$
\[
  \mathit{pre}(t) \subseteq \pl(M) \Rightarrow \mathit{post}(t) \subseteq (\pl^H - \pl(M)) \cup \mathit{pre}(t),
\] 
where $\pl(M) = \{p \in \pl^H \mid M(p) \neq \emptyset\}$.
The set of \emph{reachable markings} of $\mathcal H$ is 
\begin{align*}
\reach(\mathcal{H}) = 
\{M \mid\ & \exists\, n \ge 0\ \exists\, t_1, \ldots, t_n \in\ \tr^H\
           \exists\, v_1 \in \mathit{Val}(t_1) \ldots 
           \exists\, v_n \in \mathit{Val}(t_n):\\
         & M_0\ \fire{t_1,v_1}\ M_1\ \fire{t_2,v_2} \ldots \fire{t_n,v_n}\ M_n = M \}.
\end{align*}

\subsection{Instantiations of High-level Petri Games}
\label{subsec:inst}
For fixed parameter values,
a given high-level Petri game 
\[
 \mathcal{H} = (\plS^H,\plE^H, \tr^H, \fl^H, \init^H, \bad^H,\mathit{ty},g,e,\mathit{in})
\]
with $\pl^H = \plS^H\, \cup\, \plE^H$ can be transformed into 
a safe Petri game 
\[
  \mathcal{G} = (\plS,\plE, \tr, \fl, \init, \bad).
\]
Let $\mathcal{D} = \bigcup_{p \in \pl^H} \mathit{ty}(p)$ be the set
of all possible values that individual tokens in places $p \in \pl^H$ can take, 
and let $\mathit{Val}$ be the set of valuations assigning values $d \in \mathcal{D}$
of the right type to each variable.
The constituents of $\mathcal{G}$ are as follows:

\vspace{3mm}

\begin{itemize}
\item system places: 
    $\plS = \{ (p,d) \in \plS^H \times \mathcal{D} \mid d \in \mathit{ty}(p) \}$, \\
    
\item environment places: 
    $\plE = \{ (p,d) \in \plE^H \times \mathcal{D} \mid d \in \mathit{ty}(p) \}$, \\

\item transitions: 
    $\tr = \{ (t^H,v) \mid t^H \in \tr^H \land v \in \mathit{Val}(t^H) \land v(t^H) = true \}$, \\

\item an arc from $(p,d)$ to $(t^H,v)$ occurs in  $\fl$  if  
    $v(p,t^H) = d$ holds in $\mathcal{H}$, \\

\item an arc from $(t^H,v)$ to $(q,d)$ occurs in  $\fl$  if  
    $v(t^H,q) = d$ holds in $\mathcal{H}$, \\
    
\item initial marking: 
      $\init = \{ (p,d) \in \init^H \times \mathcal{D} \mid d \in \mathit{in}(p) \}$, \\

\item bad places: 
      $\bad = \{ (p,d) \in \bad^H \times \mathcal{D} \mid d \in \mathit{ty}(p) \}$.

\end{itemize}

\noindent The set of all places of $\mathcal{G}$ is thus given by 
\[
 \pl = \plS \cup \plE = \{ (p,d) \in \pl^H \times \mathcal{D} \mid d \in \mathit{ty}(p) \}.
\]

\begin{example}
Figure~\ref{fig:as} shows the instantiation of the alarm system for $n=2$
locations of the high-level Petri game in Fig.~\ref{fig:asHL}.
\end{example}

\subsection{Correspondence of  High-level and Low-level Petri Games}
We relate the firing behaviour of the high-level Petri game $\mathcal{H}$ to 
that of the low-level
Petri game $\mathcal{G}$ defined in Section~\ref{subsec:inst}.
To this end, we define a mapping $\rho$ from 
markings $M^H$ in $\mathcal{H}$ to sets of places in $\mathcal{G}$  as follows:
\[
 \rho(M^H) = \{ (p,d) \in \pl^H \times \mathcal{D} 
                            \mid  d \in M^H(p)\} \subseteq \pl.
\]
Note that for the initial markings $M_0$ of $\mathcal{H}$ 
and $\init$ of $\mathcal{G}$ we have
$\rho(M_0) = \init$. 
Then we can state the following correspondence that is essentially due to
\cite{jensen92}.
                        
\begin{theorem}
For all markings $M_1^H$ and $M_2^H$  of $\mathcal{H}$, all transitions $t^H \in\tr^H$,
and all valuations $v \in \mathit{Val}(t^H)$
the following properties hold:
\begin{itemize}
\item The transition $t^H$ is enabled at $M_1^H$ under $v$ in $\mathcal{H}$ 
      iff the transition $(t^H,v)$ is enabled at $\rho(M_1^H)$ in $\mathcal{G}$.

\item The firing of enabled transitions under $v$ corresponds to each other: \\
      \[
        M_1^H\ \fire{t^H,v}\ M_2^H \quad \text{iff} \quad 
        \rho(M_1^H)\ \fire{(t^H,v)}\ \rho(M_2^H).
      \]
\end{itemize}
\end{theorem}

\section{Parametric Benchmark Families}
\label{sec:benchmarkFamilies}
Using the tool {\sc Adam} \cite{fingieold15,fingiehecold17},
several benchmark families served
to demonstrate the applicability of the algorithm for solving Petri games.
With parameterized high-level Petri games these benchmark families
can now be represented concisely by one single formal object.
We exemplify this for the benchmarks \emph{Concurrent Machines} (\textbf{CM})
and \emph{Self-Reconfiguring Robots} (\textbf{SR}).
Due to the clarity of the high-level representation both families
could be optimized (in the size of the game or the functionality, respectively)
in comparison to the implemented versions of~\cite{fingieold15,fingiehecold17}.

\subsection{CM: Concurrent Machines}
\label{sec:cm}
This benchmark family models \(n\) machines of which only \(n-1\) are working correctly. 
The environment decides nondeterministically which one is defective.
The machines should process \(k\) orders and no machine is allowed to process in total more than one
order. Each order can inform itself of the defective machine and decide, with or without this information, on which machine
it would like to be processed. At the end, no order should decide for the defective or for an already used machine.
The high-level version of the benchmark family is depicted in \refFig{cmHL}.
\begin{figure}[h]
	\centering
	\scalebox{1}{
		\input{cmHL}
	}
	\caption{Parameterized high-level Petri game for the benchmark family of concurrent machines.
	There are \(k\in\N\) orders which can be processed on \(n\in\N\) machines.
	Each machine should only process one order.
	A hostile environment decides on the functionality of the machines.}
	\label{fig:cmHL}
\end{figure}
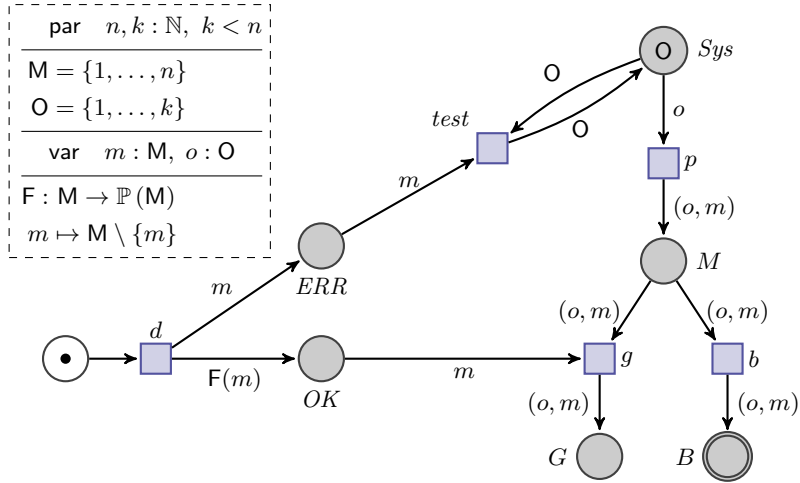

The \(n\) different machines of the family are identified by the individual tokens
in the set \({\sf M}=\{1,\ldots,n\}\).
The hostile environment decides to destroy
one of them by putting it into place \(\mathit{ERR}\) and all other but this token into place \(\mathit{OK}\)
via transition \(d\).
The \(k\) orders which should be processed by the machines are identified by the individual tokens in the set \({\sf O}=\{1,\ldots,k\}\),
which initially reside in place \(\mathit{Sys}\).
The orders can decide to first test which machine is defective (via transition \(\mathit{test}\))
and decide afterwards on which machine they want to be processed,
or choose a machine without any knowledge about the functionality of the machines (both via transition \(p\)).
A tuple \((o,m)\) residing in \(M\), for \(o\in{\sf O}\) and \(m\in{\sf M}\),
indicates that the order \(o\) should be processed by machine \(m\).
Since the place \(\mathit{OK}\) only contains one unique token for each intact machine, 
transition \(g\) can only fire at most \(|\textsf{M}|-1\) times and 
takes one of those machine identifiers of \(\mathit{OK}\) each time.
Hence, a token \((i,e)\in\textsf{O}\times\textsf{M}\) for orders \(i\)
which decide on the defective machine \(e\) or a machine \(e\) which already processed another order, 
is not moved to \(G\) but stays in \(M\).
Since we are searching for deadlock-avoiding strategies, 
this token must eventually end up in the bad place \(B\) for every strategy.

\begin{figure}[ht]
	\centering
	\scalebox{0.7}{
		\input{cm}
	}
	\caption{Instantiation of the Petri game of \refFig{cmHL} for \(|{\sf M}|=3\) and \(|{\sf O}|=2\).
	The \(k=2\) orders of this instantiation of the concurrent machines benchmark family are initially residing in \(\mathit{Sys}\)
	and \(\mathit{Sys'}\). The \(n=3\) machines are represented by the six places: \(M_i\) for the first order and \(M'_i\)
	for the second order (for \(i\in\{1,\ldots,3\})\).}
	\label{fig:cmLL}
\end{figure}
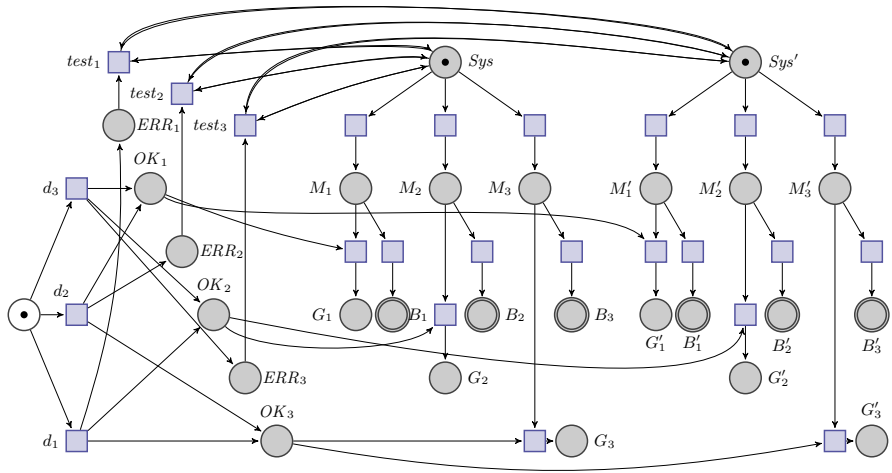
Figure~\ref{fig:cmLL} shows the instantiation of this benchmark family for three machines and two orders.
The nondeterministic destruction of machines is visualized in the left-most part,
whereas the possibilities of the two orders is depicted in the middle and the right-most part of the figure, respectively.
Each of the three machines \(m\in\{1,2,3\}\) can be functioning, i.e., a token resides in \(\mathit{OK}_i\),
or defective, i.e., a token resides in \(\mathit{ERR}_i\).
The place \(M\) collecting which order is processed on which machine of \refFig{cmHL}, as well as the corresponding 
good and bad place,
is split into \(|{\sf M}\times{\sf O}|=6\) places each.
The three undecorated copies of each of these places in the middle belong to the first
and the decorated in the most-right part of \refFig{cmLL} to the second order.
This game can be won by the system players
by first testing which of the three machines is defective,
hence knowing which two are functioning,
and afterwards unequally deciding on one of the two functioning machines.

\subsection{SR: Self-Reconfiguring Robots}
This benchmark is inspired by \cite{GudemannOrtmeierReif06}
and deals with the following scenario.
Each piece of material needs to be processed
by $n$ different tools.
There are $n$ robots having all $n$ tools
to their disposal, of which only one tool is currently used.
The environment may (repeatedly) destroy a tool on a robot $r$.
Then the robots reconfigure themselves so that $r$
uses another tool and the other robots adapt their usage
of tools accordingly.  Destructions can occur repeatedly in different phases \(p\).
We consider two safety objectives for this family of games:
\begin{enumerate}
\item \emph{No wrong tool assignment}, i.e., no robot uses a tool that is destroyed by the environment.
\item \emph{Unique tool assignment}, i.e., each tool is assigned only to a single robot.
\end{enumerate}

The high-level representation of this benchmark family is depicted in \refFig{srHL}.
\begin{figure}[t]
	\scalebox{0.95}{
		\input{srHL}
	}
	\caption{Parameterized high-level Petri game for the benchmark family of self-reconfiguring robots.
	The \(n\) robots have \(n\) tools to their disposal of which nondeterministically \(k\) tools can
	be destroyed over time.
	In this smart factory each piece of material needs to be machined by every tool.
	Every robot can use one single tool at a time and can decide on a different one 
	after the factory recognizes a defective tool on any of the robots.}
	\label{fig:srHL}
\end{figure}
The game proceeds in $k = |{\sf P}|$
phases, each one starting in the place \(\mathit{Phases}\) with type \({\sf P}\).
Initially, the game starts with phase \(1\) and for each but the last phase \(k\)
(ensured by the predicate \(p<k\))
the transition \(i_1\) puts the number of the next phase into \(\mathit{Phases}\)
and remembers the current phase in place \(S\).
For the last phase the identifier \(k\) is directly put into \(S\) via transitions \(i_2\)
and no token resides in \(\mathit{Phases}\) anymore.
Next the environment can destroy via the transition \(\mathit{des}\) 
on one robot $r$ one tool $t$ in this phase $p$
by putting the information triple $(r,t,p)$ into the system
place \(\mathit{RTP}\) and the environment token into place $W$ (for \emph{working}).
Now the system in place \(\mathit{work}\) gets active by firing transition $t_w$,
which withdraws the system token from place \(\mathit{work}\) and puts
the set of all robot identifiers equipped with the current phase \(p\) into
the system place \(\mathit{RP}\) and the environment token into 
place $C$ (for \emph{completed}).
Next the transition \(\mathit{chg}\) (for \emph{change}) is enabled.
In its preset are the places \(\mathit{RP}\) and \(\mathit{RT}\) of which the latter
contains the current assignment of tools to robots.
Here we assume w.l.o.g.\ that initially \(\mathit{RT}\) stores the assignment {\sf I}
where robot $i \in {\sf R}$ uses tool $i \in {\sf T}$.
In general, transition \(\mathit{chg}\) takes one robot identity $(r,p)$ of the current phase \(p\)
from place \(\mathit{RP}\)
and a tool assignment $(r,t)$ out of place \(\mathit{RT}\) and replaces it by
the (possibly new) assignment $(r,t')$.
The idea is that $t'$ is the tool that robot $r$ should use from now on.
The transition \(\mathit{chg}\) stores this new assignment  
by putting the triple \((r,t',p)\) into place \(\mathit{R'T'P'}\).
If $t$ is destroyed by the environment transition \(\mathit{des}\)
in any prior phase \(\widetilde{p}\) yielding $(r,t,\widetilde{p})$ in place \(\mathit{RTP}\)
then a winning strategy for the system players should choose $t \neq t'$.
Otherwise the transition \(\bot_1\) is enabled and eventually
has to fire, i.e., \(\bot_1\) puts the \emph{wrong tool assignment} $(r,t)$
into the bad place \(\mathit{Bad}_1\).
Additionally, transition \(\mathit{chg}\) stores the 
new tool assignment \((r,t')\) into place \(\mathit{check}\).
Here the \emph{unique tool assignment} property,
i.e., whether each tool is assigned only to a single robot, is checked.
The place \(\mathit{Tools}\) contains one unique identifier \(t\in{\sf T}\)
for each tool.
Every firing of transition \(c\) withdraws one of these tools.
A robot \(r\) can only reach the place \(\mathit{restart}\)
via transition \(c\) if it currently uses a tool \(t\),
i.e., the current tool assignment \((r,t)\) resides in \(\mathit{check}\),
which has not already been used by a robot already moved to \(\mathit{restart}\).
This means for two robots \(r_1\) and \(r_2\) using the same tool \(t\), i.e., 
\((r_1,t)\) and \((r_2,t)\) residing in \(\mathit{check}\), 
that one of these duplicate assignment remains in \(\mathit{check}\).
Since every winning strategy has to be deadlock-avoiding,
transition \(\bot_2\) eventually fires and puts one of the robots with the duplicate
tool assignment into the bad place \(\mathit{Bad}_2\).
When every robot uses a different tool, eventually all robots gather in place \(\mathit{restart}\)
and the transition \(\mathit{nxt}\) can enable a new phase by putting a black token
back into the environment place \(\mathit{Env}\).

\begin{figure}[H]
	\scalebox{0.77}{
		\input{sr}
	}
	\caption{Instantiation of the Petri game depicted in \refFig{srHL} for \(|{\sf R}|=|{\sf T}|=|{\sf P}|=2\).
	The \(k=2\) destruction phases are presented in the upper left and upper right part, respectively.
	The tool changing of the robots is depicted in the middle:
	in the upper left part for the first robot in the first phase,
	in the upper right part for the first robot in the second phase,
	in the lower left part for the second robot in the first phase,
	and in the lower right for the second robot in the second phase.
	The bottom of the figure presents the starting of the second phase.}
	\label{fig:sr}
\end{figure}
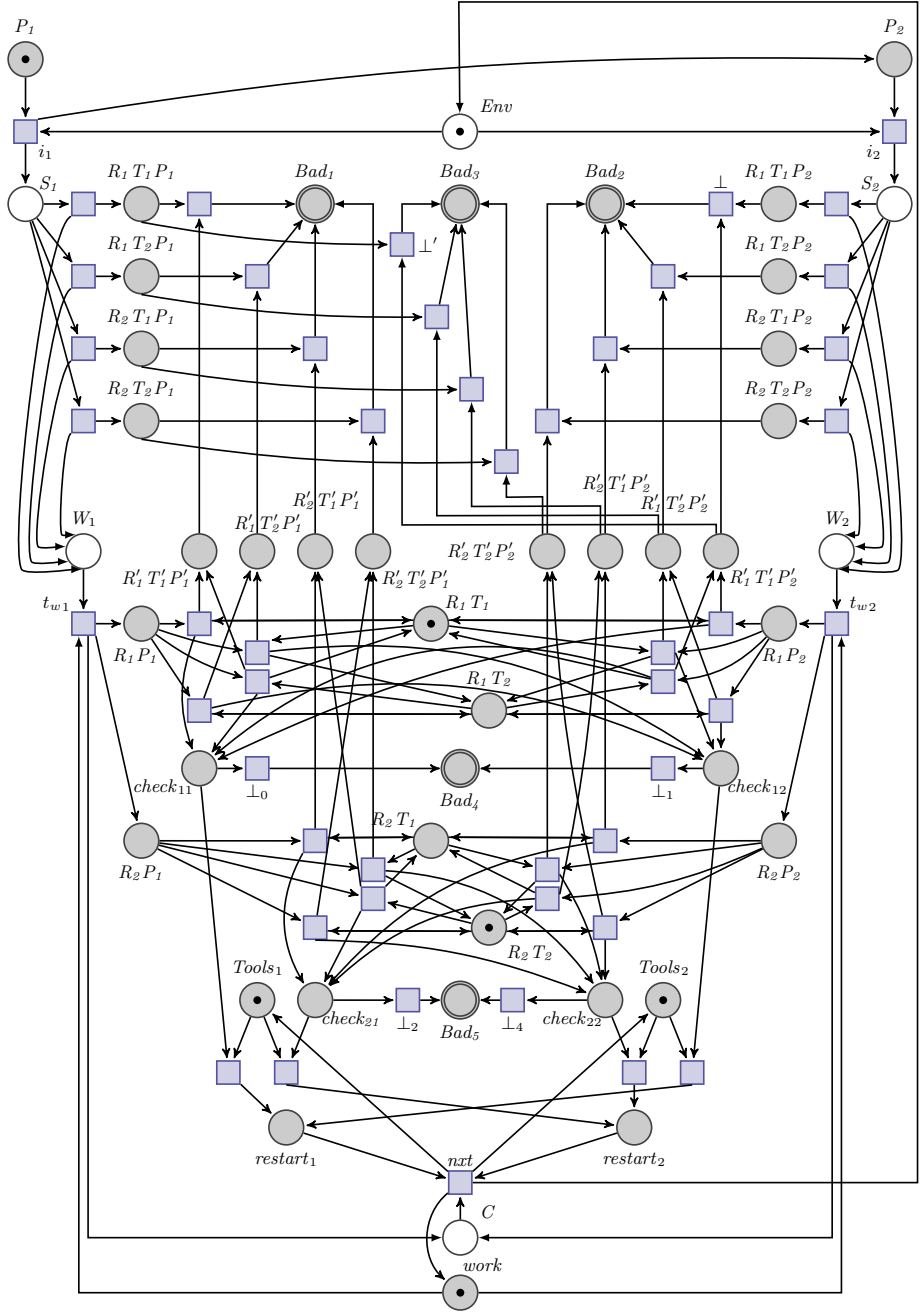

Figure~\ref{fig:sr} shows an instantiation of this high-level representation for two robots
equipped with two tools each and two destruction phases.
The first phase is presented in the left part of the figure, whereas the second part is 
depicted in the right one.
The destruction of the tools is done in the upper part of the figure.
Note that we simplified the unfolding of the high-level place \(\mathit{Bad}_1\),
for a clearer presentation.
In the middle of the figure at the first the changing of the tool for the first
and below the changing for the second robot is depicted.
The bottom of \refFig{sr} shows the starting of the second phase.

This game cannot be won by the system players,
because the environment can either decide to destroy both tools of one robot, 
say of robot \(r=1\),
or decide to destroy the same tool, say the tool \(t=1\), on each robot.
In the first case the robot has no other possibility than 
to chose an already destroyed tool, say the tool \(t=1\), in the second phase.
This enables transition \(\bot_1\) in the high-level version,
which corresponds in the low-level version to the enabledness of either \(\bot\)
or \(\bot'\), depending on the phase in which the tool \(t\) has been destroyed.
In the second case, either one of the robots decides on an already destroyed tool
(which leads us to the previous case),
or both robots decide on taking tool \(t'=2\).
This means transition \(c\) of the high-level version
can only fire for one robot, because afterwards the place \(\mathit{Tools}\) 
only contains the tool \(1\). Thus, \(\mathit{nxt}\) cannot fire
and therewith \(\bot_2\) eventually has to put the other robot into \(\mathit{Bad}_2\).
For the low-level version both robots choosing tool \(2\)
results in having a token in each of the places \(\mathit{check}_{12}\) and \(\mathit{check}_{22}\).
Hence, only one of the transitions in the postset of \(\mathit{Tools}_2\) can fire,
resulting in the eventually firing of \(\bot_1\) or \(\bot_4\), respectively.

\section{Conclusion}
\label{sec:conclusion}
We have introduced a new representation of benchmark families for Petri games.
Similarly to the advantages of high-level Petri nets versus place/transition Petri nets,
the representation captivates by its concise and complete abilities of defining the families.
The possibility to keep the expression sets parametric allows for a uniform representation of the entire family.
We have presented an instantiation technique to obtain a low-level version as standard 1-bounded Petri game for each 
element of the benchmark family.
Those Petri games can then be solved by the existing algorithms and tool.

Furthermore, we have experienced that parameterized high-level representations 
of Petri games help to understand the key ideas of benchmark families even better.
This has enabled us to improve each of the presented benchmark families compared to their original implementations
regarding their size or functionality.

Currently, the known synthesis techniques only apply for the instantiations of 
the parameterized high-level Petri games.
In future work we would like to develop algorithms to directly obtain 
parameterized distributed controllers
from such parameterized high-level representations.

\paragraph{Acknowledgement.}
We thank Wolfgang Reisig for suggesting to use high-level Petri nets
to represent families of benchmarks during a Dagstuhl Workshop
in June 2017.

\bibliographystyle{splncs04}
\bibliography{bib}
\end{document}

%% file: as.tex
\tikzstyle{light}=[opacity=0.3]
\begin{tikzpicture}[node distance=1.25cm,>=stealth',bend angle=45,auto,scale=0.8]
	\node [envplace,tokens=1]		(env)  	[label=below:$\mathit{Env}$]                                  {};
	\node [transition]			(eL)  	[left of=env, xshift=-5mm,label={[label distance=-0.5mm]below:$i_A$},label={[yshift=-2mm]above:\phantom{asdf}}] {};
	\node [transition]			(eR)  	[right of=env, xshift=5mm,label={[label distance=-0.5mm]below:$i_B$}] {};
	\node [envplace]			(epL)  	[label=left:$\mathit{CA}$, left of=eL, xshift=-5mm]                                  {};
	\node [envplace]			(epR)  	[label=right:$\mathit{CB}$, right of=eR, xshift=5mm]                                  {};
	\node [transition] 			(tA)  	[below of=epL,label={[label distance=-1mm]above right:$t_A$}] {};
	\node [sysplace]			(IA) 	[below of=tA,label=left:$\mathit{DA}$]                     {};
	\node [envplace]			(ea)  	[left of=tA,xshift=3.6mm,label=above:$\mathit{IA}$]                                  {};
	\node [sysplace]		 	(A) 	[right of=tA,label=right:$\mathit{SA}$,tokens=1]                     {};
	\node [transition]			(t2)  	[below of=IA,xshift=2mm,label=right:\(\mathit{fa}_A\)] {};
	\node [transition]			(t1)  	[left of=t2,xshift=1.6mm,label={[label distance=-1mm]left:\(\mathit{fr}_A\)}] {};
 	\node [transition]			(t3)  	[right of=t2,xshift=13mm,yshift=5mm,label=above:$\mathit{info}_B$] {};
	\node [sysplace] 			(pa) 	[below of=t2,xshift=-2mm,label=left:$\mathit{pA}$]                     {};
	\node [transition]			(tAB)  	[below left of=pa,label=right:$\mathit{aa}\phantom{b}$] {};
	\node [transition]			(tAA)  	[below right of=pa,label=left:$\mathit{ab}$] {};
	\node [sysplace]			(ab)	[below of=tAB,label=right:$\mathit{AA}$]                     {};
	\node [sysplace]			(aa) 	[below of=tAA,label=right:$\mathit{AB}$]                     {};
	\node [transition] 			(tB)  	[below of=epR,label={[label distance=-1mm]above left:$t_B$}] {};
	\node [sysplace] 			(IB) 	[below of=tB,label=right:$\mathit{DB}$]                     {};
	\node [envplace] 			(eb)  	[right of=tB,xshift=-3.6mm,label=above:$\mathit{IB}$]                                  {};
	\node [sysplace]			(B) 	[left of=tB,label=left:$\mathit{SB}$,tokens=1]                     {};
	\node [transition] 			(t2b)  	[below of=IB,xshift=-2mm,label=left:\(\mathit{fa}_B\)] {};
	\node [transition]			(t1b)  	[right of=t2b,xshift=-1.6mm,label={[label distance=-1mm]right:\(\mathit{fr}_B\)}] {};
	\node [transition] 			(t3b)  	[left of=t2b,xshift=-13mm, yshift=5mm,label=above:$\mathit{info}_A$] {};
	\node [sysplace]		 	(pb) 	[below of=t2b,xshift=2mm,label=right:$\mathit{pB}$]                     {};
	\node [transition]			(tBB)  	[below left of=pb,label=right:$\mathit{ba}$] {};
	\node [transition] 			(tBA)  	[below right of=pb,label=left:$\mathit{bb}$] {};
	\node [sysplace] 			(bb) 	[below of=tBB,label=left:$\mathit{BA}$]                     {};
	\node [sysplace] 			(ba)	[below of=tBA,label=left:$\mathit{BB}$]                     {};

	\node [transition, light] (tb1)  [right of=epR,xshift=8mm, yshift=5mm] {};
	\node [transition, light] (tb2)  [below of=tb1, yshift=5mm] {};
	\node [transition, light] (tb3)  [below of=tb2, yshift=5mm] {};
	\node [transition, light] (tb4)  [below of=tb3, yshift=5mm] {};
	\node [transition, light] (tb5)  [below of=tb4, yshift=5mm] {};
	\node [transition, light] (tb6)  [below of=tb5, yshift=5mm] {};
	\node [transition, light] (tb7)  [below of=tb6, yshift=5mm] {};
	\node [transition, light] (tb8)  [below of=tb7, yshift=5mm] {};
	\node [transition, light] (tb9)  [below of=tb8, yshift=5mm] {};
	\node [transition, light] (tb10) [below of=tb9, yshift=5mm] {};
	\node [transition, light] (tb11) [below of=tb10, yshift=5mm] {};
	\node [transition, light] (tb12) [below of=tb11, yshift=5mm] {};
	\node [sysplace,bad,light, xshift=0mm] (bad) [right of=tb6,label={[xshift=-4mm]above right:\(\mathit{Bad}\)}]                     {};
	\node [transition, light] (tg1)  [right of=bad, xshift=-4mm,yshift=25mm] {};
	\node [transition, light] (tg2)  [right of=bad, xshift=-4mm,yshift=-38mm] {};
	\node [sysplace,light] (good) at ($(tg1)!0.5!(tg2)$) [label={[xshift=2mm]below left:$\mathit{Good}$}]                     {};

	\node [right=of A,  xshift=-2mm, yshift=-4mm] (intr) {\DarkBlue{}Burglary};
	\node [below=of t3, xshift=7mm,  yshift=10mm] (com) {\color{ganttGreen}Comm.};
	\node [right=of ab, xshift=22mm, yshift=-2mm] (al) {\Red{}Alarm};

	\path[-latex, thick]
		 	(eL)  		edge [pre]                            (env)
					edge [post]                            (epL)
		 	(tA)  		edge [pre]                            (epL)
		 	  		edge [pre]                            (A)
					edge [post]                            (ea)
					edge [post]                            (IA)
			(t1)  		edge [pre]                            (IA)
					edge [post]                            (pa)
			(t2)  		edge [pre]                            (A)
					edge [post]                            (pa);
	\path[-latex, thick]
	            (t3)  		edge [pre]                            (IA)
			  		edge [pre]                            (B)
					edge [post]                            (pa)
					edge [post]                            (pb);
	\path[-latex, thick]					
			(tAB)  		edge [pre]                            (pa)
					edge [post]                            (ab)
			(tAA)  		edge [pre]                            (pa)
					edge [post]                            (aa);
	\path[->, thick]
		 	(eR)  		edge [pre]                            (env)
					edge [post]                            (epR)
		 	(tB)  		edge [pre]                            (epR)
		 	  		edge [pre]                            (B)
					edge [post]                            (eb)
					edge [post]                            (IB)
			(t1b)  		edge [pre]                            (IB)
					edge [post]                            (pb)
			(t2b)  		edge [pre]                            (B)
					edge [post]                            (pb);
	\path[-latex, thick]
			(t3b)  		edge [pre]                            (IB)
			  		edge [pre]                            (A)
					edge [post]                            (pb)
					edge [post]                            (pa);
	\path[-latex, thick]					
			(tBB)  		edge [pre]                            (pb)
					edge [post]                            (bb)
			(tBA)  		edge [pre]                            (pb)
					edge [post]                            (ba);

	\path[-latex, thick,draw=gray!50, dashed] 	
					(tb1)  
						edge [pre]                            (epL)
						edge [pre]                            (aa)
						edge [post]				(bad)
					(tb2)  
						edge [pre]                            (epL)
						edge [pre]                            (ab)
						edge [post]				(bad)
					(tb3)  
						edge [pre]                            (epL)
						edge [pre]                            (ba)
						edge [post]				(bad)
					(tb4)  
						edge [pre]                            (epL)
						edge [pre]                            (bb)
						edge [post]				(bad)

					(tb5)  
						edge [pre]                            (ea)
						edge [pre]                            (aa)
						edge [post]				(bad)
					(tb6)  
						edge [pre]                            (ea)
						edge [pre]                            (ba)
						edge [post]				(bad)

					(tb7)  
						edge [pre]                            (eb)
						edge [pre]                            (ab)
						edge [post]				(bad)
					(tb8)  
						edge [pre]                            (eb)
						edge [pre]                            (bb)
						edge [post]				(bad)

					(tb9)  
						edge [pre]                            (epR)
						edge [pre]                            (aa)
						edge [post]				(bad)
					(tb10)  
						edge [pre]                            (epR)
						edge [pre]                            (ab)
						edge [post]				(bad)
					(tb11)  
						edge [pre]                            (epR)
						edge [pre]                            (ba)
						edge [post]				(bad)
					(tb12)  
						edge [pre]                            (epR)
						edge [pre]                            (bb)
						edge [post]				(bad)
					(tg1)  
						edge [pre]                            (ea)
						edge [pre]                            (ab)
						edge [pre]                            (bb)
						edge [post]				(good)
					(tg2)  
						edge [pre]                            (eb)
						edge [pre]                            (ba)
						edge [pre]                            (aa)
						edge [post]				(good)
	;

\begin{pgfonlayer}{background}
\draw [-, rectangle,rounded corners,DarkBlue,fill=cdc_BlueL!15] ([xshift=-2mm,yshift=20mm]ea.north west) -- ([xshift=2mm,yshift=20mm]eb.north east) -- ([xshift=2mm,yshift=-5mm]eb.south east) -- ([xshift=-2mm,yshift=-5mm]ea.south west) -- cycle;
\draw [-, rectangle,rounded corners,DarkRed,fill=DarkRed!15, fill opacity=0.95] ([xshift=-2mm,yshift=-3mm]ab.south west) -- ([xshift=-2mm,yshift=35mm]ab.south west) -- ([xshift=2mm,yshift=35mm]ba.south east) -- ([xshift=2mm,yshift=-3mm]ba.south east) -- cycle;
\draw [-, rectangle,rounded corners,ganttGreen,fill=cdc_GreenL!15, fill opacity=0.45] ([xshift=-5mm,yshift=-10mm]t3.south west) -- ([xshift=-5mm,yshift=10mm]t3.north west) -- ([xshift=5mm,yshift=10mm]t3b.north east) -- ([xshift=5mm,yshift=-10mm]t3b.south east) -- cycle;
\end{pgfonlayer}
\end{tikzpicture}

%% file: asUnfolding.tex
\tikzstyle{light}=[opacity=0.3]
\begin{tikzpicture}[node distance=1.25cm,>=stealth',bend angle=45,auto,scale=0.8]
	\node [envplace,tokens=1]		 (env)  [label=below:$\mathit{Env}$]                                  {};
	\node [transition]			 (eL)  [left of=env, xshift=-5mm,label={[label distance=-0.5mm]below:$i_A$}] {};
	\node [transition]			 (eR)  [right of=env, xshift=5mm,label={[label distance=-0.5mm]below:$i_B$}] {};
	\node [envplace]			 (epL)  [label=left:$\mathit{CA}$, left of=eL, xshift=-5mm]                                  {};
	\node [envplace]			 (epR)  [label=right:$\mathit{CB}$, right of=eR, xshift=5mm]                                  {};

	\node [transition] 			(tA)  [below of=epL,label={[label distance=-1mm]above right:$t_A$}] {};
	\node [sysplace]	(IA) [below of=tA,label=left:$\mathit{DA}$]                     {};
	\node [envplace]			(ea)  [left of=tA,xshift=3.6mm,label=above:$\mathit{IA}$]                                  {};
	\node [sysplace] 	(A) [right of=tA,label=right:$\mathit{SA}$,tokens=1]                     {};
	\node [transition,draw=gray!50,fill=gray!10]			(t2)  [below of=IA,xshift=2mm,
						label=right:\color{gray!50}{\(\mathit{fa}_A\)}] {};
	\node [transition,draw=gray!50,fill=gray!10]			(t1)  [left of=t2,xshift=1.6mm,
						label=left:\color{gray!50}{\(\mathit{fr}_A\)}] {};
	\node [transition]			(t3)  [right of=t2,xshift=13mm,yshift=5mm,label=above:$\mathit{info}_B$] {};
	\node [sysplace,draw=gray!50,fill=gray!10] 	(pa) [below of=t2,xshift=-5mm,label=left:\color{gray!50}{\(\mathit{pA}_2\)}]                     {};
	\node [sysplace,draw=gray!50,fill=gray!10] (pa1) [below of=t1,xshift=-8mm,xshift=-2mm,
							label={[xshift=1mm, yshift=0mm]left:\color{gray!50}{\(\mathit{pA}_1\)}}]                     {};
	\node [sysplace] (pa2) [right of=pa,xshift=5mm,xshift=-2mm,label=left:$\mathit{pA_3}$]                     {};
	\node [sysplace] (pa3) [right of=pa2,xshift=5mm,xshift=-2mm,label=left:$\mathit{pA_4}$]                     {};
	\node [transition,draw=gray!50,fill=gray!10] (tAB)  [below of=pa,xshift=4mm] {};
	\node [transition,draw=gray!50,fill=gray!10] (tAA)  [below of=pa,xshift=-4mm] {};
	\node [transition,draw=gray!50,fill=gray!10] (tAB1)  [below of=pa1,xshift=4mm] {};
	\node [transition,draw=gray!50,fill=gray!10] (tAA1)  [below of=pa1,xshift=-4mm] {};
	\node [transition,draw=gray!50,fill=gray!10] (tAB2)  [below of=pa2,xshift=4mm] {};
	\node [transition] (tAA2)  [below of=pa2,xshift=-4mm] {};
	\node [transition] (tAB3)  [below of=pa3,xshift=4mm] {};
	\node [transition,draw=gray!50,fill=gray!10] (tAA3)  [below of=pa3,xshift=-4mm] {};
	\node [sysplace,draw=gray!50,fill=gray!10] (ab) [below of=tAB]{};
	\node [sysplace,draw=gray!50,fill=gray!10] (ab1) [below of=tAB1]{};
	\node [sysplace,draw=gray!50,fill=gray!10] (ab2) [below of=tAB2]{};
	\node [sysplace] (ab3) [below of=tAB3,label=below:\(\mathit{AB}\)]{};
	\node [sysplace,draw=gray!50,fill=gray!10] (aa) [below of=tAA]{};
	\node [sysplace,draw=gray!50,fill=gray!10] (aa1) [below of=tAA1]{};
	\node [sysplace] (aa2) [below of=tAA2,label=below:\(\mathit{AA}\)]{};
	\node [sysplace,draw=gray!50,fill=gray!10] (aa3) [below of=tAA3]{};

	\node [transition] 			(tB)  [below of=epR,label={[label distance=-1mm]above left:$t_B$}] {};
	\node [sysplace] 	(IB) [below of=tB,label=right:$\mathit{DB}$]                     {};
	\node [envplace] 			(eb)  [right of=tB,xshift=-3.6mm,label=above:$\mathit{IB}$]                                  {};
	\node [sysplace]	(B) [left of=tB,label=left:$\mathit{SB}$,tokens=1]                     {};
	\node [transition,draw=gray!50,fill=gray!10] 			(t2b)  [below of=IB,xshift=-2mm,
										label=left:\color{gray!50}{\(\mathit{fa}_B\)}] {};
	\node [transition,draw=gray!50,fill=gray!10]			(t1b)  [right of=t2b,xshift=-1.6mm,
										label=right:\color{gray!50}{\(\mathit{fr}_B\)}] {};
	\node [transition] 			(t3b)  [left of=t2b,xshift=-13mm, yshift=5mm,label=above:$\mathit{info}_A$] {};
	\node [sysplace,draw=gray!50,fill=gray!10] 	(pb) [below of=t2b,xshift=8mm,label=right:\color{gray!50}{\(\mathit{pB}_2\)}]                     {};

	\node [sysplace,draw=gray!50,fill=gray!10] (pb1) [below of=t1b,xshift=-2.5mm,xshift=15mm,label={[xshift=-1mm,yshift=0mm]right:\color{gray!50}{\(\mathit{pB}_1\)}}]                     {};
	\node [sysplace] (pb2) [left of=pb,xshift=-3mm,label=right:$\mathit{pB_3}$]                     {};
	\node [sysplace] (pb3) [left of=pb2,xshift=-3mm,label=right:$\mathit{pB_4}$]                     {};
	\node [transition,draw=gray!50,fill=gray!10] (tBB)  [below of=pb,xshift=4mm] {};
	\node [transition,draw=gray!50,fill=gray!10] (tBA)  [below of=pb,xshift=-4mm] {};
	\node [transition,draw=gray!50,fill=gray!10] (tBB1)  [below of=pb1,xshift=4mm] {};
	\node [transition,draw=gray!50,fill=gray!10] (tBA1)  [below of=pb1,xshift=-4mm] {};
	\node [transition] (tBB2)  [below of=pb2,xshift=4mm] {};
	\node [transition,draw=gray!50,fill=gray!10] (tBA2)  [below of=pb2,xshift=-4mm] {};
	\node [transition,draw=gray!50,fill=gray!10] (tBB3)  [below of=pb3,xshift=4mm] {};
	\node [transition] (tBA3)  [below of=pb3,xshift=-4mm] {};
	\node [sysplace,draw=gray!50,fill=gray!10] (bb) [below of=tBB]	                     {};
	\node [sysplace,draw=gray!50,fill=gray!10] (bb1) [below of=tBB1]	                     {};
	\node [sysplace] (bb2) [below of=tBB2,label=below:\(\mathit{BB}\)]                     {};
	\node [sysplace,draw=gray!50,fill=gray!10] (bb3) [below of=tBB3]	                     {};
	\node [sysplace,draw=gray!50,fill=gray!10] (ba) [below of=tBA]	                    {};
	\node [sysplace,draw=gray!50,fill=gray!10] (ba1) [below of=tBA1]	                    {};
	\node [sysplace,draw=gray!50,fill=gray!10] (ba2) [below of=tBA2]	                    {};
	\node [sysplace] (ba3) [below of=tBA3, label=below:\(\mathit{BA}\)]                    {};

	\node [right=of A,  xshift=-2mm, yshift=-4mm] (intr) {\DarkBlue{}Burglary};
	\node [below=of t3, xshift=7mm,  yshift=27mm] (com) {\color{ganttGreen}Comm.};
	\node [right=of ab, xshift=14.25mm, yshift=26mm] (al) {\Red{}Alarm};

	\path[-latex, thick]
		 	(eL)  		edge [pre]                            (env)
					edge [post]                            (epL)
		 	(tA)  		edge [pre]                            (epL)
		 	  		edge [pre]                            (A)
					edge [post]                            (ea)
					edge [post]                            (IA)
			(t3)  		edge [pre]                            (IA)
			  		edge [pre]                            (B)
					edge [post]                            (pa2)
					edge [post]                            (pb3)	
			(tAB3)  	edge [pre]                            (pa3)
					edge [post]                            (ab3)
			(tAA2)  	edge [pre]                            (pa2)
					edge [post]                            (aa2)		
;
	\path[->, thick, dashed, gray] 
			(t1)  		edge [pre]                            (IA)
					edge [post]                            (pa1)
			(t2)  		edge [pre]                            (A)
					edge [post]                            (pa)
			(tAB)  		edge [pre]                            (pa)
					edge [post]                            (ab)
			(tAA)  		edge [pre]                            (pa)
					edge [post]                            (aa)
			(tAB1)  		edge [pre]                            (pa1)
					edge [post]                            (ab1)
			(tAA1)  		edge [pre]                            (pa1)
					edge [post]                            (aa1)
			(tAB2)  		edge [pre]                            (pa2)
					edge [post]                            (ab2)
				
			(tAA3)  		edge [pre]                            (pa3)
					edge [post]                            (aa3)				
	;

	\path[->, thick]
		 	(eR)  		edge [pre]                            (env)
					edge [post]                            (epR)
		 	(tB)  		edge [pre]                            (epR)
		 	  		edge [pre]                            (B)
					edge [post]                            (eb)
					edge [post]                            (IB)			
			(t3b)  		edge [pre]                            (IB)
			  		edge [pre]                            (A)
					edge [post]                            (pb2)
					edge [post]                            (pa3)
			(tBB2)  	edge [pre]                            (pb2)
					edge [post]                            (bb2)	
			(tBA3)  	edge [pre]                            (pb3)
					edge [post]                            (ba3)
;
	\path[->, thick, dashed, gray] 
			(t1b)  		edge [pre]                            (IB)
					edge [post]                            (pb1)
			(t2b)  		edge [pre]                            (B)
					edge [post]                            (pb)
			(tBB)  		edge [pre]                            (pb)
					edge [post]                            (bb)
			(tBA)  		edge [pre]                            (pb)
					edge [post]                            (ba)
			(tBB1)  	edge [pre]                            (pb1)
					edge [post]                            (bb1)
			(tBA1)  	edge [pre]                            (pb1)
					edge [post]                            (ba1)
	
			(tBA2) 		edge [pre]                            (pb2)
					edge [post]                            (ba2)
			(tBB3) 		edge [pre]                            (pb3)
					edge [post]                            (bb3)
		;

\begin{pgfonlayer}{background}
\draw [-, rectangle,rounded corners,DarkBlue,fill=cdc_BlueL!15] ([xshift=-2mm,yshift=20mm]ea.north west) -- ([xshift=2mm,yshift=20mm]eb.north east) -- ([xshift=2mm,yshift=-5mm]eb.south east) -- ([xshift=-2mm,yshift=-5mm]ea.south west) -- cycle;
\draw [-, rectangle,rounded corners,DarkRed,fill=DarkRed!15, fill opacity=0.95] ([xshift=-4mm,yshift=-10mm]aa1.south west) -- ([xshift=-4mm,yshift=42mm]aa1.south west) -- ([xshift=4mm,yshift=42mm]bb1.south east) -- ([xshift=4mm,yshift=-10mm]bb1.south east) -- cycle;
\draw [-, rectangle,rounded corners,ganttGreen,fill=cdc_GreenL!15, fill opacity=0.45] ([xshift=-5mm,yshift=-8mm]t3.south west) -- ([xshift=-5mm,yshift=12mm]t3.north west) -- ([xshift=5mm,yshift=12mm]t3b.north east) -- ([xshift=5mm,yshift=-8mm]t3b.south east) -- cycle;
\end{pgfonlayer}

\end{tikzpicture}

%% file: solvingPetriGames.tex
\begin{tikzpicture}[
  ->,
thick,
  >=stealth',
  node distance=20mm,
  lbl/.style={
   align=center
  },
 ]
\tikzstyle{myarrows}=[line width=1mm,-triangle 45,postaction={line width=2mm, shorten >=2mm, -}]
\node[label=below:{\tiny{}Petri game}] 									(pg) {\includegraphics[scale=0.35]{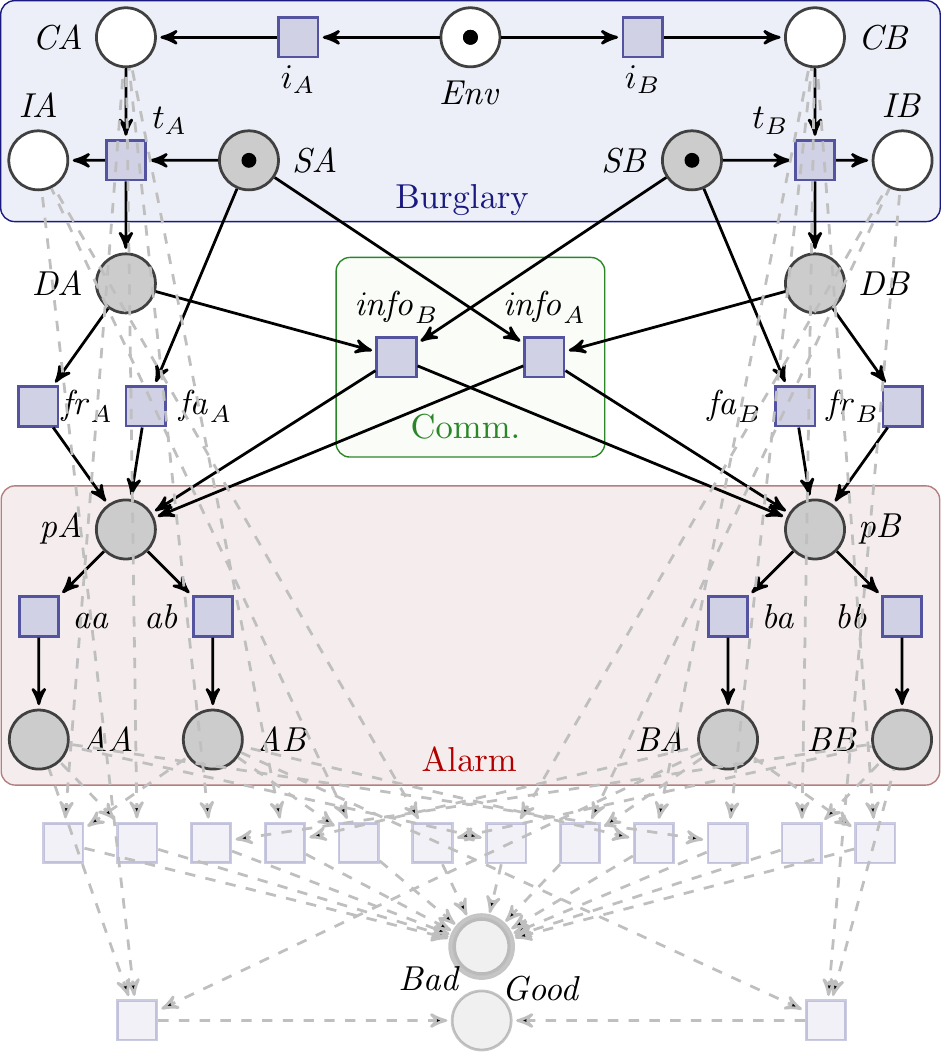}};
\node[label=below:{\tiny{}2-player game over finite graph},right=of pg, yshift=+20mm, xshift=-17.5mm] 	(gg) {\includegraphics[scale=0.2]{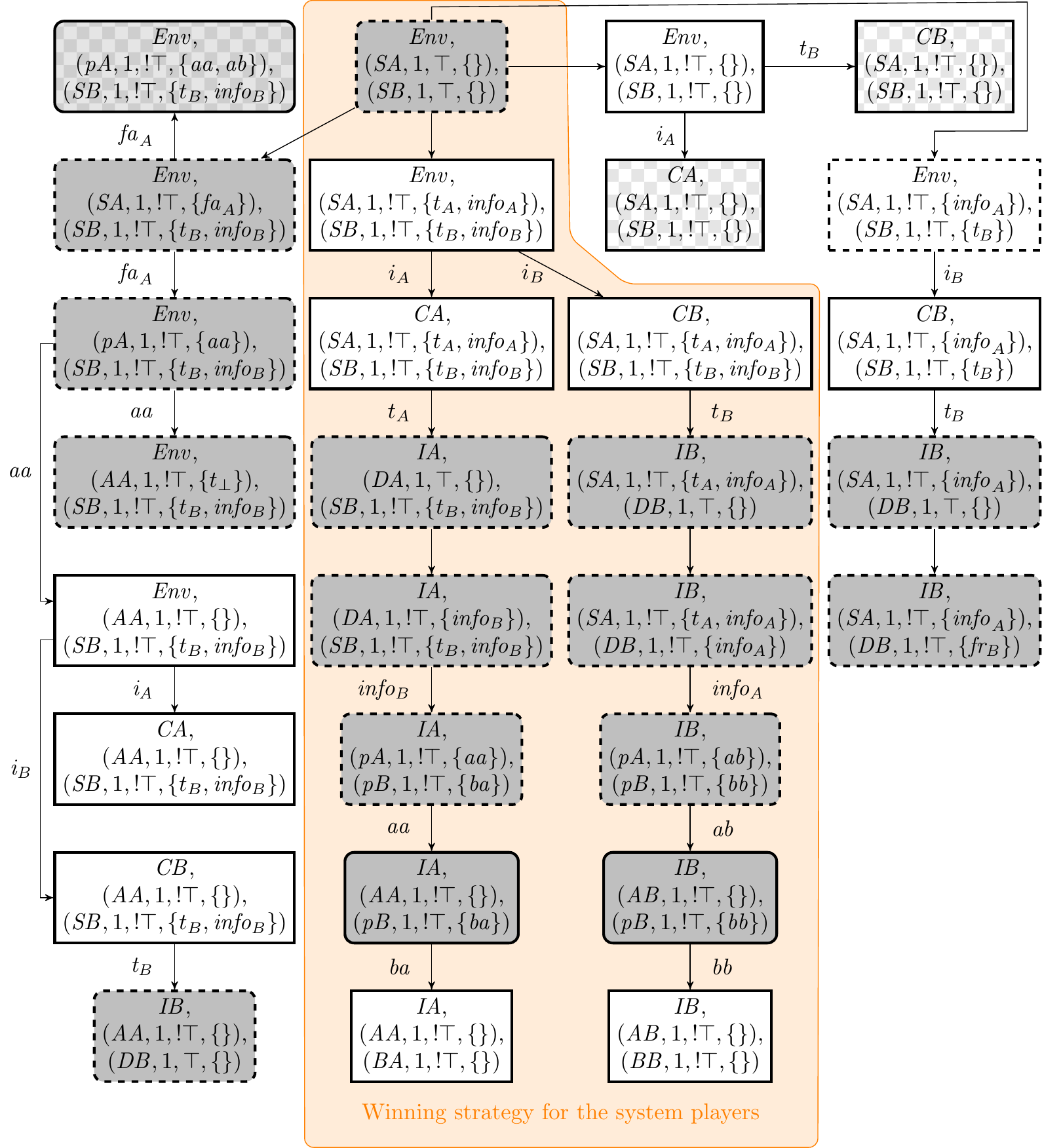}};
\node[label=below:{\tiny{}Strategy for Player 1}, right=of pg,xshift=20mm,yshift=10mm] 			(ggs) {\includegraphics[scale=0.22]{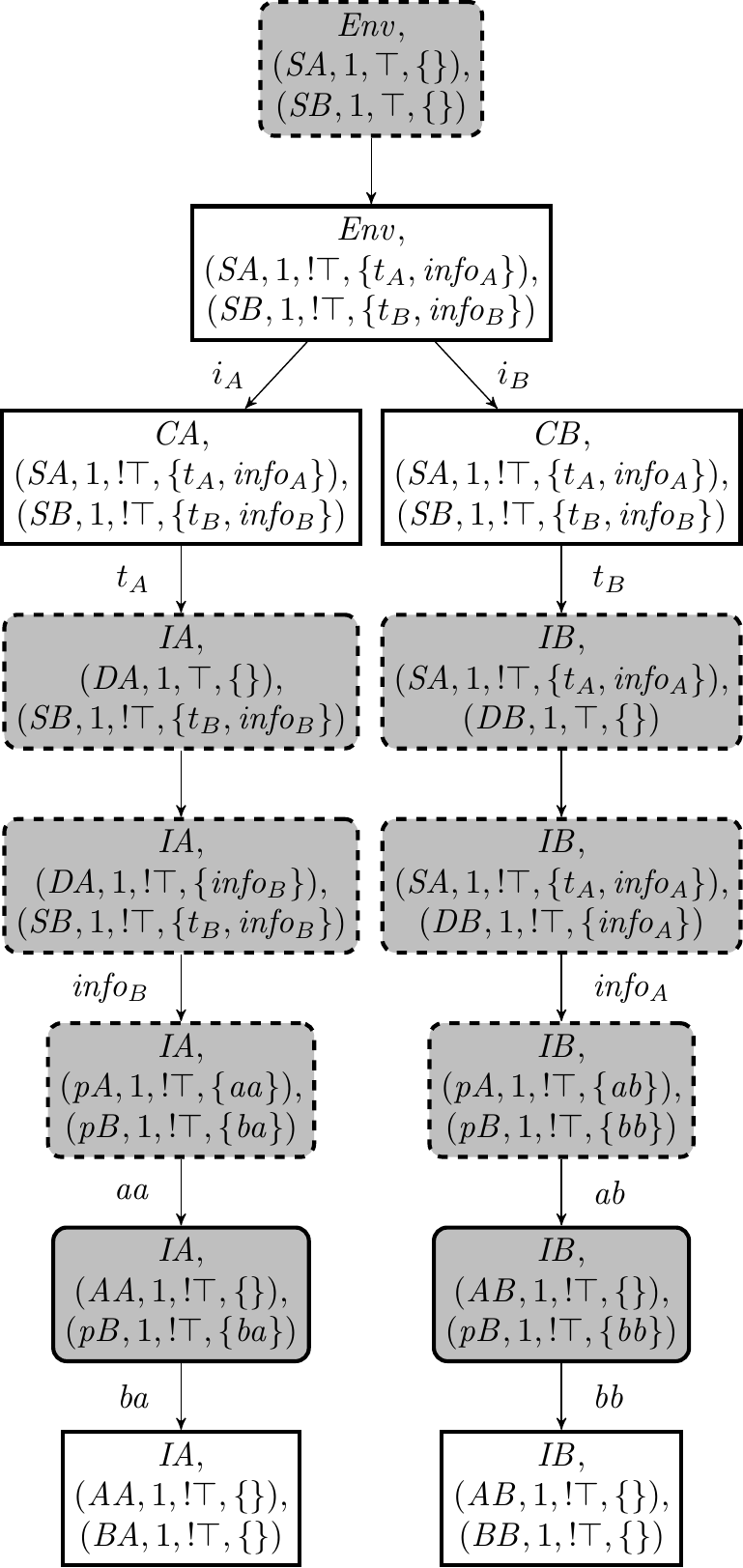}};
\node[label=below:{\tiny{}Petri game strategy for system players}, below=of gg,xshift=20mm, yshift=-5mm] (pgs) {\includegraphics[scale=0.3]{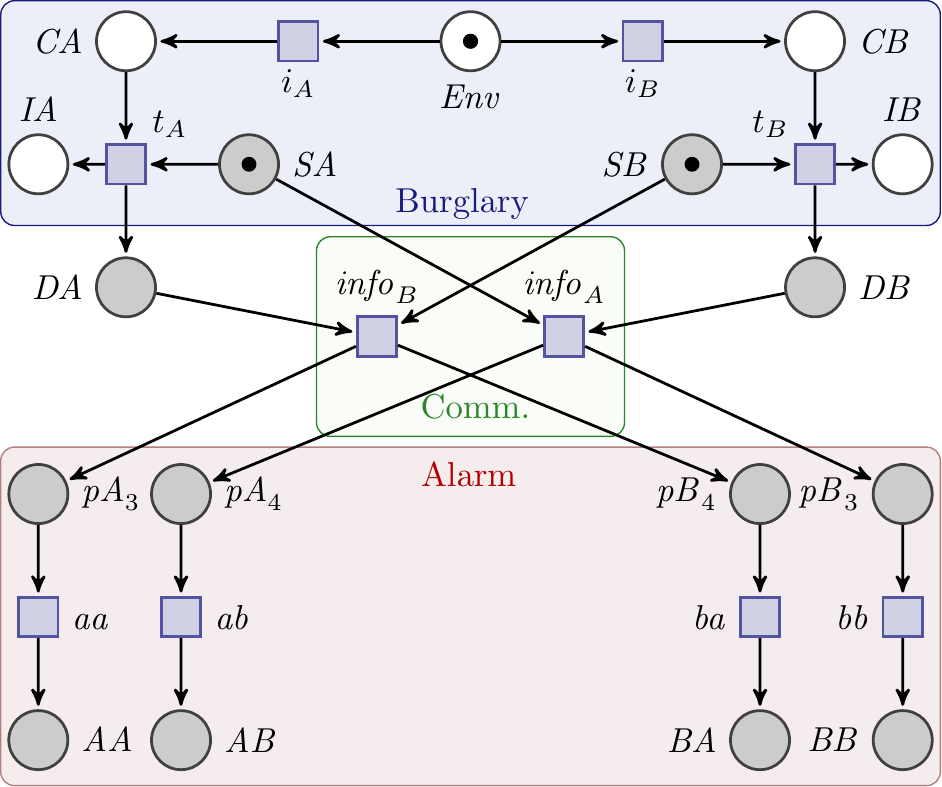}};
\node[label=below:{\tiny{}Distributed controllers}, below=of pg,xshift=4mm, yshift=15mm] 		(dc) {\includegraphics[scale=0.3]{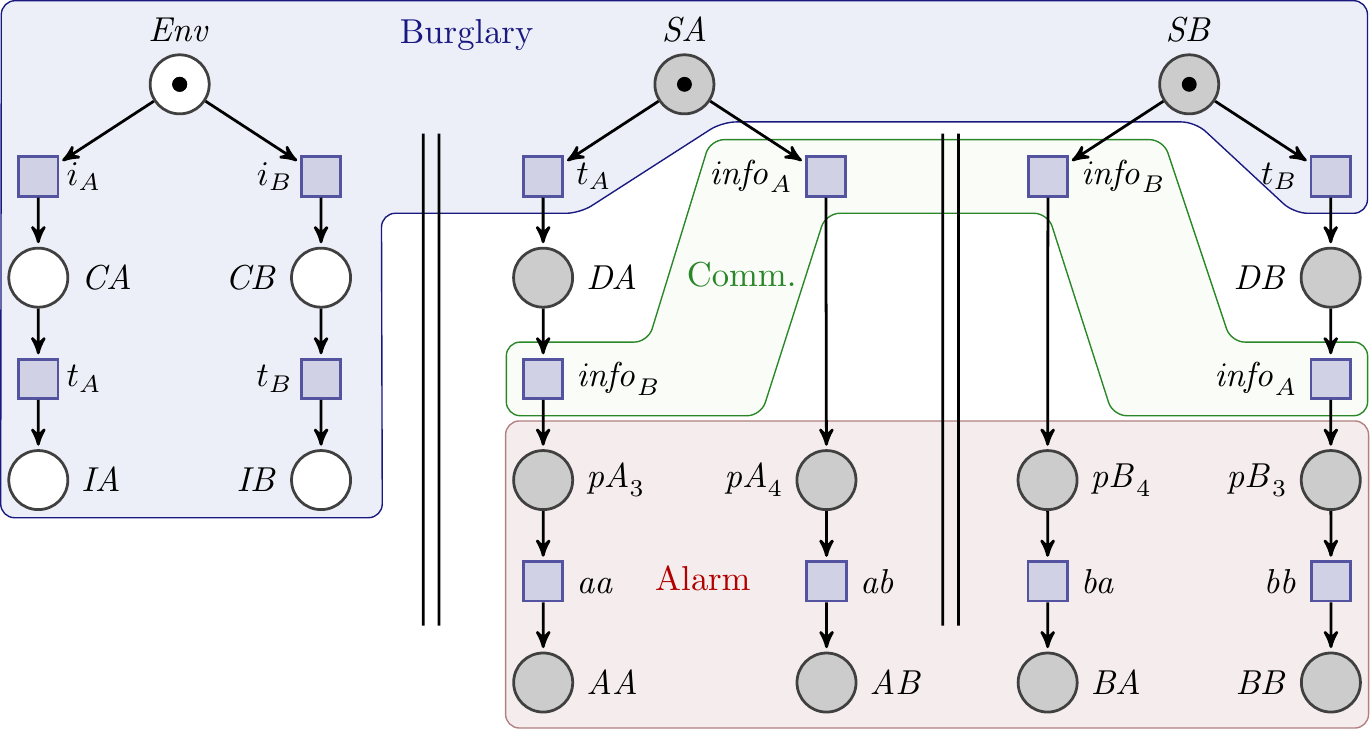}};

\path[DarkBlue,line width=1mm] 	(pg.north) 		edge[bend left=20] node[left,xshift=-2mm] {\tiny{}reduction} ([yshift=14mm]gg.west);
\path[DarkBlue,line width=1mm] ([yshift=16mm]gg.east) 	edge[bend left] node[above right, yshift=1mm,xshift=-9mm,align=center]{\tiny{}symbolic game solving} (ggs.north);
\path[DarkBlue,line width=1mm] ([yshift=-5mm, xshift=8mm]ggs.south) edge[bend left=20] node[left, yshift=2mm,xshift=0mm] {\tiny``unfolding''} ([yshift=1mm]pgs.40);
\path[DarkBlue,line width=1mm] (pgs) 			edge	 node[above, yshift=1mm] {\tiny{}distribute} (dc);
\end{tikzpicture}

%% file: asHL.tex
\begin{tikzpicture}[node distance=2cm and 1.5cm,>=stealth',bend angle=45,auto,on grid]
	\node [sysplace]			 		(sx)        	[label=above:\(Sys\)]	                         	 {\({\sf  N}\)};
	\node [transition, right =of sx,xshift=3mm]		(tx)  		[label={[label distance=-0.5mm]below:$t$}]	 {};
	\node [sysplace, right =of tx]		 		(sxx)        	[label={[label distance=-1mm,xshift=-1.5mm]above:\(D\)}]                         	 {};
	\node [transition, right =of sxx,xshift=0mm]		(info)  	[label={[label distance=-0.5mm]below:$\mathit{info}$}]	 {};
	\node [sysplace, right =of info]	 		(px)        	[label=below left:\(P\)]	                         	 {};
	\node [transition, right =of px,xshift=-2mm]		(a)		[label={[label distance=-0.5mm]below:$a$}]		 {};
	\node [sysplace, right =of a]			 	(axy)        	[label=below:\(\mathit{Alarm}\)]                    {};
	\node [transition, right =of axy,xshift=2mm]		(gt)		[label={[label distance=-0.5mm]below:$g$}]		 {};
	\node [sysplace, right =of gt,xshift=-5mm]	 	(g)        	[label=below:\(\mathit{Good}\)]		                    {};

	\node [envplace, above =of tx]			 	(lx)        	[label=above:\(C\)]	           			 {};
	\node [transition, left =of lx,xshift=5mm]		(ix)  		[label={[label distance=-0.5mm]above:$i$}]		 {};
	\node [envplace, left =of ix,tokens=1, xshift=7mm]	(env)  								{};
	\node [envplace, right =of lx]			 	(ex)        	[label=left:\(I\)]	           			 {};

	\node [transition, below =of info,xshift=0mm,yshift=5mm](fay)	  	[label={[label distance=-0.5mm]below:$\mathit{fa}$}]	 {};
	\node [transition, above =of info,xshift=0mm,yshift=-5mm](frx)  		[label={[label distance=-0.5mm]below:$\mathit{fr}$}]	 {};
	\node [transition, above =of ex,xshift=52mm]		(b1)		[label={[label distance=-1.5mm]above left:$\bot_1$}]		 {};		
	\node [transition, right =of b1,xshift=0mm]		(b2)		[label={[label distance=-1.5mm]above left:$\bot_2$}]		 {};
	\node [predicate, right =of b2, xshift=-3mm]		(prA)									 {\(b\neq x\)};
	\draw [-,dashed, thick] (b2) -- ([xshift=0.1mm]prA.west);
	\node [sysplace, bad, above =of b1] at ($(b1)!0.5!(b2)$) (bad)        	[label=right:\(\mathit{Bad}\)]                    {};

	\path[-latex, thick]
	 	(ix)	edge[pre]				node 				{}                         (env)
			edge[post] 				node[above]		 	{\(x\)}         		(lx)

	 	(tx)	edge[pre]				node[left] 			{\(x\)}                         (lx)
			edge[pre]     				node[pos=0.5,above]	 	{\(x\)} 	           	(sx)
			edge[post]      			node[above]	 	{\(x\)}		            	(ex)
			edge[post]      			node[above]		 	{\(x\)}		            	(sxx)

	 	(frx)	edge[bend right=25, pre]		node[above] 		{\(x\)}                 (sxx)
			edge[bend left, post]	  		node[above]	 	{\(x\)}	            	(px)

	 	(fay)	edge[bend left=30, pre]			node[below] 		{\(y\)}                 (sx)
			edge[bend right, post]  		node[below]	 	{\(y\)}	           	(px)

	 	(info)	edge[pre]				node[above,pos=0.7] 		{\(x\)}                         (sxx)
			edge[pre, bend left=30]			node[below,pos=0.5] 		{\({\sf  F}(x)\)}                      (sx)
			edge[post]  				node[above]	 	{\({\sf  N}\)}            	(px)

	 	(a)	edge[pre]				node[above] 		{\(z\)}                         (px)
			edge[post]  				node[above]	 	{\((z,v)\)}	            	(axy)

		(gt.north)edge[pre,bend right=15, looseness=1.2]	node[below] 		{\(x\)}                         (ex)
		(gt)	edge[pre]				node[above] 		{\({\sf  G}(x)\)}            (axy)
			edge[post]  				node[above]	 	{\(\)}		            	(g)

	 	(b1)	edge[pre]				node[left] 		{\((a,b)\)}                     (axy)
		(b1)	edge[pre,bend right=10]			node[above]		{\(x\)} 	                (lx)
			edge[post]      			node[left]	 	{\(\)}		            	(bad)

	 	(b2)	edge[pre]				node[right] 		{\((a,b)\)}                     (axy)
			edge[pre, bend left=10]			node[above]	 		{\(x\)} 	                (ex)
			edge[post]      			node[right]	 	{\(\)}		            	(bad)
	;

	\legend[0mm,xshift=30mm,yshift=-10mm]{north west}{
	    & {\sf par}\quad n: \mathbb{N} \\[1mm]\hline\\[-4mm]
		{\sf  N}&=\{1,\ldots,\, n\}\\[1mm]\hline\\[-4mm]
		& {\sf var} \quad x,y,z,v,a,b:{\sf  N}\\[1mm]\hline\\[-4mm]
		{\sf  F}: &\ {\sf  N}\rightarrow\pom{\sf  N}
		,\; x\mapsto{\sf  N}\setminus\{x\}\\\hline\\[-4mm]
		{\sf  G}: &\ {\sf  N}\rightarrow\pom{{\sf  N}\times{\sf  N}}
		,\; x\mapsto \{(z,x) \mid z\in{\sf  N}\}\\[1mm]
	}

	\node [above=of env,  xshift=4mm, yshift=-12mm] (intr) {\DarkBlue{}Burglary};
	\node [below=of info, xshift=0mm,  yshift=12.5mm] (com) {\color{ganttGreen}Comm.};
	\node [below=of a, xshift=0mm, yshift=12.5mm] (al) {\Red{}Alarm};

\begin{pgfonlayer}{background}
\draw [-, rectangle,rounded corners,DarkBlue,fill=cdc_BlueL!15]
 ([xshift=-2mm,yshift=-3mm]sx.south west) 
-- ++(0mm,37mm)
-- ++(27mm,0mm)
-- ++(0mm,-6.5mm)
-- ++(15mm,0mm)
-- ++(0mm,-10mm)
-- ++(-14mm,0mm)
-- ++(0mm,-20.5mm)
-- cycle;
\draw [-, rectangle,rounded corners,DarkRed,fill=DarkRed!15, fill opacity=0.95] ([xshift=-4mm,yshift=3mm]px.north west) -- ([xshift=3.5mm,yshift=3mm]axy.north east) -- ([xshift=3.5mm,yshift=-8mm]axy.south east) -- ([xshift=-4mm,yshift=-8mm]px.south west) -- cycle;
\draw [-, rectangle,rounded corners,ganttGreen,fill=cdc_GreenL!15, fill opacity=0.45] ([xshift=-3.5mm,yshift=3mm]info.north west) -- ([xshift=3.5mm,yshift=3mm]info.north east) -- ([xshift=3.5mm,yshift=-8mm]info.south east) -- ([xshift=-3.5mm,yshift=-8mm]info.south west) -- cycle;
\end{pgfonlayer}
\end{tikzpicture}

%% file: cmHL.tex
\begin{tikzpicture}[node distance=1.5cm and 1.7cm,>=stealth',bend angle=45,auto,on grid]
	\node [envplace, tokens =1] 					(env)  		                            					{};
	\node [transition, xshift=-5mm,right =of env]			(a)  		[label={[label distance=-0.5mm]above:$d$}]		 {};
	\node [sysplace, right =of a,xshift=5mm]	 	(ax)        	[label=below:\(\mathit{OK}\)]	           			 {};
	\node [sysplace, above =of ax]		 		(tx)        	[label=below:\(\mathit{ERR}\)]	                         	 {};
	\node [transition, right =of ax,xshift=20mm]		(gx)  		[label={[label distance=-0.5mm]right:$g$}]		 {};
	\node [transition, right =of gx]			(bx)  		[label={[label distance=-0.5mm]right:$b$}]		 {};
	\node [sysplace, below =of gx,yshift=2mm]		 		(good)        	[label=left:\(G\)]	                         	 {};
	\node [sysplace, bad, below =of bx,yshift=2mm]	 		(bad)        	[label=left:\(B\)]	                         	 {};
	\node [sysplace, above =of gx,yshift=-2mm] at ($(gx)!0.5!(bx)$)	(mx)        	[label=right:\(M\)]	                         	 {};
	\node [transition, above =of mx,yshift=-2mm]			(wx)  		[label={[label distance=-0.5mm]right:$p$}]		 {};
	\node [sysplace, above =of wx]		 		(sys)        	[label=right:\(\mathit{Sys}\)]	                         {\({\sf O}\)};
	\node [transition] at ($(tx)!0.5!(sys)$)		(test)		[label={[label distance=-0.5mm]above left:$\mathit{test} $}] {};

	\path[-latex, thick]
	 	(a)	edge[pre]	node 			{}                         (env)
			edge[post]      node[below]	 	{\({\sf F}(m)\)}            (ax)
			edge[post]      node[above,xshift=-2mm]	{\(m\)}            (tx)

	 	(test)	edge[pre]	node[above] 		{\(m\)}                         (tx)
			edge[pre, bend left=10]       node		 	{\({\sf O}\)}            	(sys)
			edge[post, bend right=10]      node[below]		 	{\({\sf O}\)}            	(sys)

	 	(wx)	edge[pre]	node[right] 		{\(o\)}                         (sys)
			edge[post]      node		 	{\((o,m)\)}	            	(mx)

	 	(gx)	edge[pre]	node[left] 		{\((o,m)\)}                     (mx)
			edge[pre]	node	 		{\(m\)}                     (ax)
			edge[post]      node[left]	 	{\((o,m)\)}	            	(good)

	 	(bx)	edge[pre]	node[right] 		{\((o,m)\)}                         (mx)
			edge[post]      node		 	{\((o,m)\)}	            	(bad)
	;

	\legend[25mm,xshift=-12mm,yshift=-14mm]{north west}{
	 & {\sf par} \quad n,k:\mathbb{N},\  k < n\\[1mm]
	 \hline\\[-4mm]
	  {\sf M}&=\{1,\ldots,n\}\\
	  {\sf O}&=\{1,\ldots,k\}\\[1mm]
	 \hline\\[-4mm]
	 & {\sf var} \quad m: {\sf M},\; o: {\sf O} \quad \\[1mm]
	 \hline\\[-4mm]
	 {\sf F}:&\ {\sf M}\rightarrow\pom{\sf M}\\
	 m & \mapsto{\sf M}\setminus\{m\} 	
	 }
	 
\end{tikzpicture}

%% file: cm.tex
\begin{tikzpicture}[node distance=12mm,>=stealth',bend angle=15,auto, remember picture,scale=0.5]
% system 1
	\node [sysplace,tokens=1] (sys)  [label=right:\(\mathit{Sys}\)]                                  {};
	\node [transition] (tm1)  [below of=sys] {};
	\node [transition] (tm0)  [left of=tm1, xshift=-5mm] {};
	\node [transition] (tm2)  [right of=tm1, xshift=+5mm] {};
	\node [sysplace] (m0)  [below of=tm0, label=left:\(M_1\)]                                  {};
	\node [sysplace] (m1)  [below of=tm1, label=left:\(M_2\)]                                  {};
	\node [sysplace] (m2)  [below of=tm2, label=left:\(M_3\)]                                  {};
	\node [transition] (tg0)  [below of=m0] {};
	\node [transition] (tb0)  [right of=tg0,xshift=-5mm] {};
	\node [transition] (tb1)  [below of=m1, xshift=+7mm] {};
	\node [transition] (tb2)  [below of=m2, xshift=+7mm] {};
	\node [sysplace] (g0)  [below of=tg0, label=left:\(G_1\)]                                  {};
	\node [sysplace,bad] (b0)  [below of=tb0, label={[xshift=-1.4mm]right:\(B_1\)}]                                  {};
	\node [sysplace,bad] (b1)  [below of=tb1, label=right:\(B_2\)]                                  {};
	\node [sysplace,bad] (b2)  [below of=tb2, label=right:\(B_3\)]                                  {};
	\node [transition] (tg1)  [below of=m1, yshift=-12mm] {};
	\node [sysplace] (g1)  [below of=tg1, label=right:\(G_2\)]                                  {};
	\node [transition] (tg2)  [below of=m2, yshift=-36mm] {};
	\node [sysplace] (g2)  [right of=tg2, xshift=-5mm, label=right:\(G_3\)]                                  {};

% system2
	\node [sysplace,tokens=1] (sys1)  [right of=sys,xshift=45mm,label=right:\(\mathit{Sys}'\)]                                  {};
	\node [transition] (tm11)  [below of=sys1] {};
	\node [transition] (tm01)  [left of=tm11, xshift=-5mm] {};
	\node [transition] (tm21)  [right of=tm11, xshift=+5mm] {};
	\node [sysplace] (m01)  [below of=tm01, label=left:\(M_1'\)]                                  {};
	\node [sysplace] (m11)  [below of=tm11, label=left:\(M_2'\)]                                  {};
	\node [sysplace] (m21)  [below of=tm21, label=left:\(M_3'\)]                                  {};
	\node [transition] (tg01)  [below of=m01] {};
	\node [transition] (tb01)  [right of=tg01,xshift=-5mm] {};
	\node [transition] (tb11)  [below of=m11, xshift=+7mm] {};
	\node [transition] (tb21)  [below of=m21, xshift=+7mm] {};
	\node [sysplace] (g01)  [below of=tg01, label={[yshift=0.5mm]below:\(G_1'\)}]                                  {};
	\node [sysplace,bad] (b01)  [below of=tb01, label={[yshift=0.5mm]below:\(B_1'\)}]                                  {};
	\node [sysplace,bad] (b11)  [below of=tb11, label=below:\(B_2'\)]                                  {};
	\node [sysplace,bad] (b21)  [below of=tb21, label=below:\(B_3'\)]                                  {};
	\node [transition] (tg11)  [below of=m11, yshift=-12mm] {};
	\node [sysplace] (g11)  [below of=tg11, label=right:\(G_2'\)]                                  {};
	\node [transition] (tg21)  [below of=m21, yshift=-36mm] {};
	\node [sysplace] (g21)  [right of=tg21,xshift=-5mm, label=above:\(G_3'\)]                                  {};

% env
	\node [transition] (test0)  [left of=sys, xshift=-50mm,label=left:\(\mathit{test}_1\)] {};
	\node [transition] (test1)  [right of=test0, yshift=-6mm,label=left:\(\mathit{test}_2\)] {};
	\node [transition] (test2)  [right of=test1, yshift=-6mm,label=left:\(\mathit{test}_3\)] {};
	\node [sysplace] (ta0)  [below of=test0, label={[xshift=-1mm]right:\(\mathit{ERR}_1\)}]                                  {};
	\node [sysplace] (a0)  [below of=ta0, xshift=6mm, label=above:\(\mathit{OK}_1\)]                                  {};
	\node [sysplace] (ta1)  [below of=a0, xshift=6mm, label={[xshift=-1mm]right:\(\mathit{ERR}_2\)}]                                  {};
	\node [sysplace] (a1)  [below of=ta1, xshift=6mm, label=above:\(\mathit{OK}_2\)]                                  {};
	\node [sysplace] (ta2)  [below of=a1, xshift=6mm, label={[xshift=-1mm]right:\(\mathit{ERR}_3\)}]                                  {};
	\node [sysplace] (a2)  [below of=ta2, xshift=6mm, label=above:\(\mathit{OK}_3\)]                                  {};
	\node [transition] (t2)  [left of=a0, xshift=-2mm, label=left:\(d_3\)] {};
	\node [transition] (t1)  [below of=t2, yshift=-12mm, label={[xshift=2mm]above left:\(d_2\)}] {};
	\node [transition] (t0)  [below of=t1, yshift=-12mm, label=left:\(d_1\)] {};
	\node [envplace, tokens=1] (env)  [left of=t1, xshift=2mm]                                  {};

	\draw[->] 	(t0)  		edge [pre]                            (env)
					edge [post]                            (a1)
					edge [post, bend right=10]                            (ta0)
					edge [post]                            (a2)
			(t1)  		edge [pre]                            (env)
					edge [post]                            (a0)
					edge [post]                            (ta1)
					edge [post]                            (a2)
			(t2)  		edge [pre]                            (env)
					edge [post]                            (a1)
					edge [post]                            (ta2)
					edge [post]                            (a0)
			(test0)		edge [pre]                            (ta0)
					edge [pre, bend left=5, in=140, looseness=0.2]                   (sys)
					edge [post, bend left=5, in=140, looseness=0.2]           (sys)
					edge [pre, bend left=80, in=140, looseness=0.3]                   (sys1)
					edge [post, bend left=80, in=140, looseness=0.3]                   (sys1)
			(test1)		edge [pre]                            (ta1)
					edge [pre, bend left=7, in=160, looseness=0.4]                   (sys)
					edge [post, bend left=7, in=160, looseness=0.4]       	(sys)
					edge [pre, bend left=60, in=160, looseness=0.5]                   (sys1)
					edge [post, bend left=60, in=160, looseness=0.5]                   (sys1)
			(test2)		edge [pre]                            (ta2)
					edge [pre, bend left=5]                   (sys)
					edge [post, bend left=5]             (sys)
					edge [pre, bend left=80, in=170,looseness=0.65]                   (sys1)
					edge [post, bend left=80, in=170,looseness=0.65]                   (sys1)
			(tm0)  		edge [pre]                            (sys)
					edge [post]                            (m0)
			(tm1)  		edge [pre]                            (sys)
					edge [post]                            (m1)
			(tm2)  		edge [pre]                            (sys)
					edge [post]                            (m2)
			(tg0)  		edge [pre]                            (m0)
					edge [pre, bend left=5]                            (a0)
					edge [post]                            (g0)
			(tg1)  		edge [pre]                            (m1)
					edge [pre, bend left=45, looseness=.5]                            (a1)
					edge [post]                            (g1)
			(tg2)  		edge [pre]                            (m2)
					edge [pre]                            (a2)
					edge [post]                            (g2)
			(tb0)  		edge [pre]                            (m0)
					edge [post]                            (b0)
			(tb1)  		edge [pre]                            (m1)
					edge [post]                            (b1)
			(tb2)  		edge [pre]                            (m2)
					edge [post]                            (b2)

			(tm01)  		edge [pre]                            (sys1)
					edge [post]                            (m01)
			(tm11)  		edge [pre]                            (sys1)
					edge [post]                            (m11)
			(tm21)  		edge [pre]                            (sys1)
					edge [post]                            (m21)
			(tg01)  		edge [pre]                            (m01)
					edge [pre, bend right=-25, out=-40, looseness=.4]                            (a0)
					edge [post]                            (g01)
			(tg11)  		edge [pre]                            (m11)
					edge [pre, bend right=100, in=170, out=80, looseness=.45]                            (a1)
					edge [post]                            (g11)
			(tg21)  		edge [pre]                            (m21)
					edge [pre, bend left=10]                            (a2)
					edge [post]                            (g21)
			(tb01)  		edge [pre]                            (m01)
					edge [post]                            (b01)
			(tb11)  		edge [pre]                            (m11)
					edge [post]                            (b11)
			(tb21)  		edge [pre]                            (m21)
					edge [post]                            (b21)
;
\end{tikzpicture}

%% file: srHL.tex
\begin{tikzpicture}[node distance=1.2cm and 2cm,>=stealth',bend angle=45,auto,on grid]
	\node [sysplace,tokens=1]		 		(prog)        	[label=above:\(\mathit{work}\)]	                         	 {};
	\node [transition, right =of prog,xshift=-1cm]		(tx)  		[label={[label distance=-0.5mm]below:$t_w$}]	 {};
	\node [sysplace, right =of tx]		 		(sxx)        	[label={[label distance=-1mm,xshift=-0mm]above:\(\mathit{RP}\)}]                         	 {};
	\node [transition, right =of sxx,xshift=0mm]		(info)  	[label={[label distance=-1.5mm,xshift=2mm]below left:$\mathit{chg}$}]	 {};
	\node [sysplace, right =of info]	 		(px)        	[label=above:\(\mathit{RT}\)]	               	 {\({\sf I}\)};
	\node [sysplace, right =of px,xshift=-10mm]		(check)        	[label={[label distance=-.5mm,xshift=-0.4mm]below:\(\mathit{check}\)}]	               	 {\(\)};
	\node [sysplace, right =of check,xshift=5mm]		(axy) 	 	[label=above:\(\mathit{restart}\)]                		 {};
	\node [transition] at ($(check)!0.5!(axy)$)		(cc)		[label={[label distance=-.5mm]below:\(\mathit{c}\)}] {};
	\node [sysplace, above =of cc]				(tools)        	[label=above:\(\mathit{Tools}\)]                		 {\({\sf T}\)};
	\node [transition, right =of axy,xshift=-7mm]		(gt)		[label={[label distance=-1.5mm,yshift=.5mm,xshift=-2mm]above right:\(\mathit{nxt}\)}] {};

	\node [envplace, above =of tx]			 	(eP)        	[label=left:\(W\)]	           			 {};
	\node [transition, above =of eP]		 	(dP)        	[label=left:\(\mathit{des}\)]	           			 {};
	\node [sysplace, right =of dP]			 	(deR)        	[label=above:\(\mathit{RTP}\)]	           			 {};
	\node [sysplace, above =of info]		 	(deR2)        	[label=left:\(\mathit{R'T'P'}\)]	           			 {};
	\node [envplace, above =of dP]			 	(lx)        	[label=left:\(S\)]	           			 {};
	\node [transition, above=of lx, xshift=5mm]		(ix)  		[label={[label distance=-0.5mm]left:\(i_1\)}]		 {};
	\node [transition, above=of lx, xshift=-5mm]		(ix2)  		[label={[label distance=-0.5mm]left:\(i_2\)}]		 {};
	\node [predicate, right =of ix, xshift=-10mm]		(prB)									 {\(p< k\)};
	\draw [-,dashed, thick] (ix) -- ([xshift=0.1mm]prB.west);
	\node [sysplace, above =of ix,yshift=5mm]		(env)  		[label=above:$\mathit{Phases}$, label=right:\({\sf P}\)]		{\(1\)};
	\node [envplace, above =of ix2,yshift=5mm,tokens=1] 	(envBuf)        [label=left:\(\mathit{Env}\)]               	 {};
	\node [envplace, below =of sxx]			 	(ex)        	[label=left:\(C\)]	           			 {};

	\node [transition, above =of deR2]		(b2)		[label={[label distance=-1.5mm]below right:$\bot_1$}]		 {};
	\node [predicate, above =of b2, yshift=-3mm]		(prA)									 {\(p\leq p'\)};
	\draw [-,dashed, thick] (b2) -- ([xshift=0.1mm]prA.south);
	\node [transition, above =of check]		(b1)		[label={[label distance=0mm]left:$\bot_2$}]		 {};
	\node [sysplace, bad, above =of b1] 		 	(bad)        	[label=above:\(\mathit{Bad_2}\)]                    {};
	\node [sysplace, bad, above =of px,yshift=12mm] 	 	(bad2)        	[label=above:\(\mathit{Bad_1}\)]                    {};	

	\path[-latex, thick]
	 	(ix)	edge[pre]			node[right]			{\(p\)}                         (env)
	 	(ix) edge[pre]					              					       (envBuf)
		(ix)	edge[post] 				node[left]		 	{\(p\)}         		(lx)
			edge[post, bend right] 			node[right]		 	{\(p+1\)}         		(env)

		(ix2)	edge[pre]			node[below, pos=0.95] 				{\(k\)}                         (env)
	 		edge[pre]					              					       (envBuf)
			edge[post] 				node[left]		 	{\(k\)}         		(lx)

	 	(tx)	edge[pre]				node[left] 			{\(p\)}                         (eP)
			edge[pre]     				node[pos=0.5,above]	 	{\(\)}  	           	(prog)
			edge[post]      			node[below left]	 	{\(\)}		            	(ex)
			edge[post]      			node[below,pos=0.7]		 	{\({\sf F}(p)\)}           	(sxx)

		(dP)	edge[pre]				node[left] 			{\(p\)}                         (lx)
			edge[post]      			node[above]		 	{\((r,t,p)\)}	            	(deR)
			edge[post]      			node[left]		 	{\(p\)}           	(eP)

	 	(info)	edge[pre]				node[below] 		{\((r,p)\)}                         (sxx)
			edge[pre, bend left]  			node[below]	 	{\((r,t)\)}            	(px)
			edge[post, bend right]  		node[above]	 	{\((r,t')\)}            	(px)
			edge[post]  				node[left]	 	{\((r,t',p)\)}            	(deR2)
		(info.south)	edge[post, bend right]  	node[below]	 	{\((r,t')\)}            	(check)

		(cc)	edge[pre]				node[below] 		{\((r,t)\)}                         (check)
			edge[pre]     				node[right]	 	{\(t\)}  	           	(tools)
			edge[post]      			node[below]	 	{\(r\)}		            	(axy)

		(gt.south west)edge[pre,bend left=15, looseness=1.2]	node[above] 		{\(\)}                         (ex)
		(gt)	edge[pre]				node[below,pos=0.7] 		{\({\sf R}\)}            (axy)
		(gt)	edge[post, bend right=20]				node[above] 		{\({\sf T}\)}            (tools)

	 	(b2)	edge[pre]				node[above] 		{\((r,t,p)\)}                     (deR)
			edge[pre]				node[left]	 	{\((r,t,p')\)} 	                (deR2)
			edge[post]      			node[above]	 	{\((r,p')\)}		            	(bad2)
		(b1) 	edge[pre]			node[right]			{\((r,t)\)}			(check)
			edge[post]				node[right]		{\(r\)}				(bad)
	;
	\draw[-latex, thick] (gt.north) -- ++(0,36mm) -- ++(-86mm,0) -- ++(0, 38mm) -- ++(-27mm, 0) -- (envBuf.north);
	\draw[-latex, thick] (gt.south) -- ++(0,-16mm) -- ++(-118mm,0) -- (prog.south);

	\legend[-35mm,yshift=-13mm,xshift=-13mm]{north east}{
	    & {\sf par} \quad n,k:\mathbb{N}\\[1mm]\hline\\[-4mm]
		{\sf R} & =\{1,\ldots, n\}\\[-1mm]
		{\sf T} & =\{1,\ldots, n\}\\[-1mm]
		{\sf P} & =\{1,\ldots, k\}\\
		{\sf I} & =\{(i,i)\in{\sf R}\times{\sf T} \mid i\in\{1,\ldots,n\}\}\\[1mm]\hline\\[-4mm]
		& {\sf var} \quad r:{\sf R},\; t,t':{\sf T},\; p,p':{\sf P}\quad\\[1mm]
	 \hline\\[-4mm]
	 {\sf F}:&\ {\sf P}\rightarrow\pom{\sf R\times \sf P}\\
	 p & \mapsto\{(r,p)\with r\in{\sf R}\} 
	}
\end{tikzpicture}

%% file: sr.tex
\begin{tikzpicture}[node distance=1.25cm and 1cm,>=stealth',bend angle=45,auto,scale=0.8, on grid]
%%%%%%% phase 1 
	\node [sysplace, tokens=1]					(p0) 	 	[label=above:$\mathit{P_1}$]    		 	{};
	\node [transition, below=of p0, xshift=0mm] 			(t0)  		[label={[label distance=-1.5mm]below right:$i_1$}]	{};
	\node [envplace, below=of t0]					(s0)  		[label={[label distance=-1.5mm]above right:$\mathit{S_1}$}]		     	{};
	\node [transition, right=of s0]		 			(d000) 		[label={[label distance=-0.5mm]below:$ $}]	{};
	\node [sysplace,right=of d000]					(r0d0p0)  	[label=above:$\mathit{R_1T_1P_1}$]    		 	{};
	\node [transition, below=of d000] 				(d010) 		[label={[label distance=-0.5mm]below:$ $}]	{};
	\node [sysplace,right=of d010]					(r0d1p0)  	[label=above:$\mathit{R_1T_2P_1}$]    		 	{};
	\node [transition, below=of d010]	 			(d100) 		[label={[label distance=-0.5mm]below:$ $}]	{};
	\node [sysplace,right=of d100]					(r1d0p0)  	[label=above:$\mathit{R_2T_1P_1}$]    		 	{};
	\node [transition, below=of d100]	 			(d110) 		[label={[label distance=-0.5mm]below:$ $}]	{};
	\node [sysplace,right=of d110]					(r1d1p0)  	[label=above:$\mathit{R_2T_2P_1}$]    		 	{};

	\node [envplace, below=of d110, yshift=-10mm]			(w0) 		[label=above:$\mathit{W}_1$]		     	{};
	\node [transition, below=of w0, xshift=0mm]	 		(tw0)  		[label={[label distance=-1.5mm]above left:${t_w}_1$}] {};
	\node [sysplace,right=of w0, xshift=10mm]			(r0d0p0_)  	[label={[xshift=2mm,yshift=+0.2mm]below left:$\mathit{R_1'T_1'P_1'}$}]   {};
	\node [sysplace,right=of r0d0p0_]				(r0d1p0_)  	[label={[xshift=2mm,yshift=-1mm]above:$\mathit{R_1'T_2'P_1'}$}]   {};
	\node [sysplace,right=of r0d1p0_]				(r1d0p0_)  	[label={[xshift=2mm,yshift=3mm]above:$\mathit{R_2'T_1'P_1'}$}]    		{};
	\node [sysplace,right=of r1d0p0_]				(r1d1p0_)  	[label={[xshift=-2mm,yshift=0.2mm]below right:$\mathit{R_2'T_2'P_1'}$}]	 	{};
	\node [sysplace,right=of tw0, yshift=0mm]			(r0p0)  		[label={[xshift=-1mm]below:$\mathit{R_1P_1}$}]   		 	{};
	\node [sysplace,below=of r0p0, yshift=-25mm]			(r1p0)  		[label=below:$\mathit{R_2P_1}$]    		 	{};
%%% bad
	\node [transition, xshift=0mm]	at (r0d0p0 -| r0d0p0_)		(tb1)		[label={[label distance=-0.5mm]below:$ $}] {};
	\node [transition, xshift=0mm]	at (r0d1p0 -| r0d1p0_) 		(tb2)		[label={[label distance=-0.5mm]below:$ $}] {};
	\node [transition, xshift=0mm]	at (r1d0p0 -| r1d0p0_) 		(tb3)		[label={[label distance=-0.5mm]below:$ $}] {};
	\node [transition, xshift=0mm]	at (r1d1p0 -| r1d1p0_) 		(tb4)		[label={[label distance=-0.5mm]below:$ $}] {};
	\node [sysplace,bad] 		at (r0d0p0 -| r1d0p0_)		(bad1) 	 	[label=above:$\mathit{Bad_1}$]    		 	{};
%%% robot1
	\node [sysplace,right=of r1d1p0_,tokens=1,yshift=-12.5mm]	(r0t0) 		[label={[label distance=-1.5mm]above right:$\mathit{R_1T_1}$}]    		 	{};
	\node [sysplace,right=of r0t0,yshift=-15mm]			(r0t1) 		[label=above:$\mathit{R_1T_2}$]    		 	{};
	\node [transition]		 at (r0t0 -| r0d0p0_)		(r0a)  		[label={[label distance=-0.5mm]below:$ $}] {};
	\node [transition, yshift=-5mm]	 at (r0t0 -| r0d1p0_)		(r0b)  		[label={[label distance=-0.5mm]below:$ $}] {};
	\node [transition, xshift=0mm]	 at (r0t1 -| r0d0p0_)		(r0c)  		[label={[label distance=-0.5mm]below:$ $}] {};
	\node [transition, yshift=5mm]	 at (r0t1 -| r0d1p0_)		(r0d)  		[label={[label distance=-0.5mm]below:$ $}] {};

	\node [sysplace,below=of r0d0p0_,yshift=-25mm]			(check00)	[label={[label distance=-2mm,xshift=1mm]below left:$\mathit{check}_{11}$}] 	{};
	\node [transition, right=of check00, xshift=0mm] 		(badC00)	[label={[label distance=-.5mm]below:${\bot}_0$}] {};
%%% robot 2
	\node [sysplace]	 	at (r1p0 -| r0t0)		(r1t0) 		[label={[label distance=-1.5mm]above left:$\mathit{R_2T_1}$}]    		 	{};
	\node [sysplace, right=of r1t0, tokens=1,yshift=-15mm]		(r1t1) 		[label={[label distance=-0.5mm]below right:$\mathit{R_2T_2}$}]    		 	{};
	\node [transition]		 at (r1t0 -| r1d0p0_)		(r1a)  		[label={[label distance=-0.5mm]below:$ $}] {};
	\node [transition, yshift=-5mm]	 at (r1t0 -| r1d1p0_)		(r1b)  		[label={[label distance=-0.5mm]below:$ $}] {};
	\node [transition, xshift=0mm]	 at (r1t1 -| r1d0p0_)		(r1c)  		[label={[label distance=-0.5mm]below:$ $}] {};
	\node [transition, yshift=5mm]	 at (r1t1 -| r1d1p0_)		(r1d)  		[label={[label distance=-0.5mm]below:$ $}] {};

	\node [sysplace,below=of r1c,yshift=0mm]			(check10)	[label={[label distance=-2mm,xshift=-1mm]below right:$\mathit{check_{21}}$}]		 	{};
	\node [transition, right=of check10, xshift=6mm] 		(badC10)	[label={[label distance=-.5mm]below:${\bot}_2$}] {};

%%%%%%%% phase 2
	\node [sysplace, right=of p0, xshift=140mm]			(p1) 	 	[label=above:$\mathit{P_2}$]    		 	{};
	\node [transition, below=of p1, xshift=0mm] 			(t1)  		[label={[label distance=-1.5mm]below left:$i_2$}]	{};
	\node [envplace, below=of t1]					(s1)  		[label={[label distance=-1.5mm]above left:$\mathit{S_2}$}]		     	{};
	\node [transition, left=of s1]		 			(d001) 		[label={[label distance=-0.5mm]below:$ $}]	{};
	\node [sysplace,left=of d001]					(r0d0p1)  	[label=above:$\mathit{R_1T_1P_2}$]    		 	{};
	\node [transition, below=of d001] 				(d011) 		[label={[label distance=-0.5mm]below:$ $}]	{};
	\node [sysplace,left=of d011]					(r0d1p1)  	[label=above:$\mathit{R_1T_2P_2}$]    		 	{};
	\node [transition, below=of d011]	 			(d101) 		[label={[label distance=-0.5mm]below:$ $}]	{};
	\node [sysplace,left=of d101]					(r1d0p1)  	[label=above:$\mathit{R_2T_1P_2}$]    		 	{};
	\node [transition, below=of d101]	 			(d111) 		[label={[label distance=-0.5mm]below:$ $}]	{};
	\node [sysplace,left=of d111]					(r1d1p1)  	[label=above:$\mathit{R_2T_2P_2}$]    		 	{};

	\node [envplace, tokens=1] at ($(t0)!.5!(t1)$)			(env)  		[label=above right:$\mathit{Env}$]		     	{};
	\node [envplace, below=of d111, yshift=-10mm]			(w1) 		[label=above:$\mathit{W}_2$]		     	{};
	\node [transition, below=of w1, xshift=0mm]	 		(tw1)  		[label={[label distance=-1.5mm]above right:${t_w}_2$}] {};
	\node [sysplace,left=of w1, xshift=-10mm]			(r0d0p1_)  	[label={[xshift=-2mm,yshift=0.5mm]below right:$\mathit{R_1'T_1'P_2'}$}] {};
	\node [sysplace,left=of r0d0p1_]				(r0d1p1_)  	[label={[xshift=2mm,yshift=2.5mm]above:$\mathit{R_1'T_2'P_2'}$}]	 	{};
	\node [sysplace,left=of r0d1p1_]				(r1d0p1_)  	[label={[xshift=2mm,yshift=6mm]above:$\mathit{R_2'T_1'P_2'}$}]	 	{};
	\node [sysplace,left=of r1d0p1_]				(r1d1p1_)  	[label={[xshift=-1mm]left:$\mathit{R_2'T_2'P_2'}$}]    		 	{};
	\node [sysplace,left=of tw1, yshift=0mm]			(r0p1)  		[label={[xshift=1mm]below:$\mathit{R_1P_2}$}]    		 	{};
	\node [sysplace,below=of r0p1, yshift=-25mm]			(r1p1)  		[label=below:$\mathit{R_2P_2}$]    		 	{};
%%% bad
	\node [transition, xshift=0mm]	at (r0d0p1 -| r0d0p1_)		(tb5)		[label={[label distance=-0.5mm]above:$\bot$}] {};
	\node [transition, xshift=0mm]	at (r0d1p1 -| r0d1p1_) 		(tb6)		[label={[label distance=-0.5mm]below:$ $}] {};
	\node [transition, xshift=0mm]	at (r1d0p1 -| r1d0p1_) 		(tb7)		[label={[label distance=-0.5mm]below:$ $}] {};
	\node [transition, xshift=0mm]	at (r1d1p1 -| r1d1p1_) 		(tb8)		[label={[label distance=-0.5mm]below:$ $}] {};
	\node [sysplace,bad] 		at (r0d0p1 -| r1d0p1_)	 	(bad2) 	 	[label=above:$\mathit{Bad_2}$]    		 	{};
	\node [sysplace,bad] 		at ($(bad1)!.5!(bad2)$)	 	(bad3) 	 	[label=above:$\mathit{Bad_3}$]    		 	{};
	\node [transition, right =of tb1, xshift=25mm, yshift=-7mm]	(tbb0)		[label={[label distance=-0.5mm]right:$\bot'$}] {};
	\node [transition, right =of tb2, xshift=21mm, yshift=-7mm]	(tbb1)		[label={[label distance=-0.5mm]below:$ $}] {};
	\node [transition, right =of tb3, xshift=17mm, yshift=-7mm]	(tbb2)		[label={[label distance=-0.5mm]below:$ $}] {};
	\node [transition, right =of tb4, xshift=13mm, yshift=-7mm]	(tbb3)		[label={[label distance=-0.5mm]below:$ $}] {};

%%% robot1
	\node [transition]		 at (r0t0 -| r0d0p1_)		(r0ap2)  		[label={[label distance=-0.5mm]below:$ $}] {};
	\node [transition, yshift=-5mm]	 at (r0t0 -| r0d1p1_)		(r0bp2)  		[label={[label distance=-0.5mm]below:$ $}] {};
	\node [transition, xshift=0mm]	 at (r0t1 -| r0d0p1_)		(r0cp2)  		[label={[label distance=-0.5mm]below:$ $}] {};
	\node [transition, yshift=5mm]	 at (r0t1 -| r0d1p1_)		(r0dp2)  		[label={[label distance=-0.5mm]below:$ $}] {};
	
	\node [sysplace,below=of r0d0p1_,yshift=-25mm]			(check01)	[label={[xshift=-1mm,label distance=-2mm]below right:$\mathit{check}_{12}$}] 	{};
	\node [transition, left=of check01, xshift=0mm] 		(badC01)	[label={[label distance=-.5mm]below:${\bot}_1$}] {};
	\node [sysplace, bad] at ($(badC00)!.5!(badC01)$)		(bad4)  	[label=below:$\mathit{Bad_4}$]		     	{};

%%% robot2
	\node [transition]		 at (r1t0 -| r1d0p1_)		(r1ap2)  		[label={[label distance=-0.5mm]below:$ $}] {};
	\node [transition, yshift=-5mm]	 at (r1t0 -| r1d1p1_)		(r1bp2)  		[label={[label distance=-0.5mm]below:$ $}] {};
	\node [transition, xshift=0mm]	 at (r1t1 -| r1d0p1_)		(r1cp2)  		[label={[label distance=-0.5mm]below:$ $}] {};
	\node [transition, yshift=5mm]	 at (r1t1 -| r1d1p1_)		(r1dp2)  		[label={[label distance=-0.5mm]below:$ $}] {};

	\node [sysplace,below=of r1cp2]					(check11)	[label={[label distance=-2mm,xshift=1.5mm]below left:$\mathit{check}_{22}$}] {};
	\node [transition, left=of check11, xshift=-6mm] 		(badC11)	[label={[label distance=-.5mm]below:${\bot}_4$}] {};
	\node [sysplace, bad] at ($(check10)!.5!(check11)$)		(bad5)  	[label=below:$\mathit{Bad_5}$]		     	{};

%%%%%%%%%%%%%%% check 
	\node [sysplace,left=of check10,tokens=1]			(tools1)	[label=above:$\mathit{Tools}_1$]    		 	{};
	\node [transition, below=of tools1] at ($(tools1)!.5!(check10)$)(c1)  	[label=below:$\mathit{}$]		     	{};
	\node [transition, left=of c1] 				(c2)  	[label=below:$\mathit{}$]		     	{};
	\node [sysplace,below=of c1,yshift=3mm]					(res1)	[label=below:$\mathit{restart}_1$]    		 	{};

	\node [sysplace,right=of check11,tokens=1]			(tools2)	[label=above:$\mathit{Tools}_2$]    		 	{};
	\node [transition, below=of tools2] at ($(tools2)!.5!(check11)$)(c3)  	[label=below:$\mathit{}$]		     	{};
	\node [transition, right=of c3] 				(c4)  	[label=below:$\mathit{}$]		     	{};
	\node [sysplace,below=of c3,yshift=3mm]					(res2)	[label=below:$\mathit{restart}_2$]    		 	{};

	\node [transition, below=of res1,yshift=3mm]	at ($(res1)!.5!(res2)$)	(nxt)		[label={[label distance=-0.5mm]above:$\mathit{nxt}$}] {};
	\node [envplace, below=of nxt,yshift=3mm] 			(c) 		[label=above right:$\mathit{C}$]		     	{};
	\node [sysplace, tokens=1, below=of c,yshift=3mm] 	(work) 		[label={[label distance=-1.5mm, xshift=-2mm,yshift=1.5mm]above right:$\mathit{work}$}]		
		     	{};

%%%%%%%%%%%%%%%%%%%%%%%%%%%%%%%%%%%%%%%%%%% PATHS
%%%%% phase1
	\path[-latex, thick]
		 	(t0) 		edge [pre]                            (env)
					edge [pre]                            (p0)
					edge [post]                            (s0)
			(t0.north east)	edge [post,bend left=5]                            (p1)
		 	(d000) 		edge [pre]                            (s0)
					edge [post]                            (r0d0p0)
		 	(d010) 		edge [pre]                            (s0)
					edge [post]                            (r0d1p0)
		 	(d100) 		edge [pre]                            (s0)
					edge [post]                            (r1d0p0)
		 	(d110) 		edge [pre]                            (s0)
					edge [post]                            (r1d1p0);
	\draw [-{latex}, thick] (d000.south west)  to[out=200,in=120, looseness=.2] ($(w0.265) - (13mm,0)$)  to[out=300,in=180, looseness=.3] (w0.265);
	\draw [-{latex}, thick] (d010.south west)  to[out=200,in=120, looseness=.2] ($(w0.215) - (8mm,0)$)  to[out=300,in=180, looseness=.3] (w0.215);
	\draw [-{latex}, thick] (d100.south west)  to[out=200,in=120, looseness=.2] ($(w0.165) - (6mm,0)$)  to[out=300,in=180, looseness=.3] (w0.165);
	\draw [-{latex}, thick] (d110.south west)  to[out=200,in=120, looseness=.2] ($(w0.115) - (3mm,0)$)  to[out=300,in=180, looseness=.3] (w0.115);
	\path[-latex,thick]
			(tw0) 		edge [pre] 				(w0)
					edge [post]				(r0p0)
			(tw0.south east)		edge [post]				(r1p0);

	\draw [-{latex}, thick] ([xshift=1mm]tw0.south)  -- ([xshift=1mm]c -| tw0)  -- (c.west);
	\draw [-{latex}, thick] (work.west)  -- ([xshift=-1mm]work -| tw0)  -- ([xshift=-1mm]tw0.south);
%% bad
	\path[-latex, thick]
		 	(tb1) 		edge [pre]                            (r0d0p0)
					edge [pre,bend left=0]                (r0d0p0_)
					edge [post]                            (bad1)
		 	(tb2) 		edge [pre]                            (r0d1p0)
					edge [pre]                            (r0d1p0_)
					edge [post]                            (bad1)
		 	(tb3) 		edge [pre]		                 (r1d0p0)
					edge [post]                            (bad1)
			(tb3) 		edge [pre, bend right=0]                 (r1d0p0_)
		 	(tb4) 		edge [pre]                            (r1d1p0)
					edge [pre]                            (r1d1p0_);
	\draw [-{latex}, thick] (tb4)  -- ([xshift=0mm]bad1 -| tb4)  -- ([xshift=0mm]bad1.east);
%% robot1
	\path[-latex, thick]
		 	(r0a) 		edge [pre, bend left=0]                            (r0p0)
			([yshift=-2mm]r0a.north east) edge [pre]                  ([yshift=-2mm,xshift=-1mm]r0t0.north west)
					edge [post]                 ([yshift=-2mm,xshift=-1mm]r0t0.north west)
			(r0a)		edge [post]                            (r0d0p0_)
					edge [post,bend right=20]                            (check00)

		 	(r0b) 		edge [pre,bend left=5]                   (r0p0)
			(r0b.north east)edge [pre]                            (r0t0)
			(r0b)		edge [post]                            (r0t1.150)
					edge [post]                            (r0d1p0_)
					edge [post, bend left=20]                            (check01)

		 	(r0c) 		edge [pre, bend left=0]                            (r0p0)
			([yshift=2mm]r0c.south east)		edge [pre]     ([yshift=2mm,xshift=-1mm]r0t1.south west)
					edge [post]                            ([yshift=2mm,xshift=-1mm]r0t1.south west)
			(r0c)		edge [post, bend left=0]                            (r0d1p0_)
					edge [post, bend left=19]                            (check01)
		 	(r0d) 		edge [pre, bend left=10]                            (r0p0)
					edge [pre]                            (r0t1)
					edge [post,bend left=0]                            (r0t0)
			(r0d.north west)	edge [post, bend left=0]                            (r0d0p0_)
			(r0d.south)		edge [post]                            (check00)
			(badC00)	edge [pre]				(check00)
					edge [post]				(bad4);
%% robot2
	\path[-latex, thick]
		 	(r1a) 		edge [pre]                            (r1p0)
			([yshift=-2mm]r1a.north east) edge [pre]                  ([yshift=-2mm,xshift=-1mm]r1t0.north west)
					edge [post]                 ([yshift=-2mm,xshift=-1mm]r1t0.north west)
			(r1a)		edge [post]                            (r1d0p0_)
			(r1a.south west)edge [post,bend right=30]                            (check10)
		 	(r1b) 		edge [pre]                            (r1p0)
					edge [pre]                            (r1t0)
					edge [post]                            (r1t1)
					edge [post]                            (r1d1p0_)
					edge [post, bend left=25]                     (check11)
		 	(r1c) 		edge [pre]                            (r1p0)
			([yshift=2mm]r1c.south east)		edge [pre]                            ([yshift=2mm,xshift=-1mm]r1t1.south west)
					edge [post]                            ([yshift=2mm,xshift=-1mm]r1t1.south west)
			(r1c)		edge [post, bend left=0]                            (r1d1p0_)
			(r1c.south)		edge [post, bend left=10]                            (check11)
		 	(r1d) 		edge [pre]                            (r1p0)
					edge [pre]                            (r1t1)
					edge [post]                            (r1t0)
			(r1d.north west)	edge [post, bend left=0]                            (r1d0p0_)
			(r1d.south west)	edge [post]                            (check10)
			(badC10)	edge [pre]				(check10)
					edge [post]				(bad5);

%%%%%%%%%% phase2
	\path[-latex, thick]
		 	(t1) 		edge [pre]                            (env)
					edge [pre]                            (p1)
					edge [post]                            (s1)
		 	(d001) 		edge [pre]                            (s1)
					edge [post]                            (r0d0p1)
		 	(d011) 		edge [pre]                            (s1)
					edge [post]                            (r0d1p1)
		 	(d101) 		edge [pre]                            (s1)
					edge [post]                            (r1d0p1)
		 	(d111) 		edge [pre]                            (s1)
					edge [post]                            (r1d1p1);
	\draw [-{latex}, thick] (d001.south east)  to[out=-20,in=60, looseness=.2] ($(w1.-85) + (13mm,0)$)  to[out=-120,in=0, looseness=.3] (w1.-85);
	\draw [-{latex}, thick] (d011.south east)  to[out=-20,in=60, looseness=.2] ($(w1.-35) + (8mm,0)$)  to[out=-120,in=0, looseness=.3] (w1.-35);
	\draw [-{latex}, thick] (d101.south east)  to[out=-20,in=60, looseness=.2] ($(w1.15) + (6mm,0)$)  to[out=-120,in=0, looseness=.3] (w1.15);
	\draw [-{latex}, thick] (d111.south east)  to[out=-20,in=60, looseness=.2] ($(w1.65) + (3mm,0)$)  to[out=-120,in=0, looseness=.3] (w1.65);
	\path[-latex,thick]
			(tw1) 		edge [pre] 				(w1)
					edge [post]				(r0p1)
			(tw1.south west)		edge [post]				(r1p1);
	\draw [-{latex}, thick] ([xshift=-1mm]tw1.south)  -- ([xshift=-1mm]c -| tw1)  -- (c.east);
	\draw [-{latex}, thick] (work.east)  -- ([xshift=1mm]work -| tw1)  -- ([xshift=1mm]tw1.south);
 
%% bad
	\path[-latex, thick]
		 	(tb5) 		edge [pre]                            (r0d0p1)
					edge [pre,bend left=0]                (r0d0p1_)
					edge [post]                            (bad2)
		 	(tb6) 		edge [pre]                            (r0d1p1)
					edge [pre]                            (r0d1p1_)
					edge [post]                            (bad2)
		 	(tb7) 		edge [pre]		                 (r1d0p1)
					edge [post]                            (bad2)
			(tb7) 		edge [pre, bend right=0]                 (r1d0p1_)
		 	(tb8) 		edge [pre]                            (r1d1p1)
					edge [pre]                            (r1d1p1_);
	\draw [-{latex}, thick] (tb8)  -- ([xshift=0mm]bad2 -| tb8)  -- ([xshift=0mm]bad2.west);
	\path[-latex, thick]
		 	(tbb0) 		edge [pre, bend left=5]               (r0d0p0.south);
	\draw [-{latex}, thick] (tbb0)  -- ([xshift=0mm]bad3 -| tbb0)  -- ([xshift=0mm]bad3.west);
	\draw [-{latex}, thick] (r0d0p1_.105)  -- ++(0,2mm)  -- ([yshift=5.7mm] r0d0p1_-| tbb0)  -- ([xshift=0mm]tbb0.south);
	\path[-latex, thick]
		 	(tbb1) 		edge [pre, bend left=5]               (r0d1p0.south)
					edge [post]                            (bad3);
	\draw [-{latex}, thick] (r0d1p1_.105)  -- ++(0,4mm)  -- ([yshift=7.7mm] r0d1p1_-| tbb1)  -- ([xshift=0mm]tbb1.south);
	\path[-latex, thick]
		 	(tbb2) 		edge [pre, bend left=5]               (r1d0p0.south)
					edge [post]                            (bad3);
	\draw [-{latex}, thick] (r1d0p1_.105)  -- ++(0,6mm)  -- ([yshift=9.7mm] r1d0p1_-| tbb2)  -- ([xshift=0mm]tbb2.south);
	\path[-latex, thick]
		 	(tbb3) 		edge [pre, bend left=5]               (r1d1p0.south);
	\draw [-{latex}, thick] (tbb3)  -- ([xshift=0mm]bad3 -| tbb3)  -- ([xshift=0mm]bad3.east);
	\draw [-{latex}, thick] (r1d1p1_.105)  -- ++(0,8mm)  -- ([yshift=11.7mm] r1d1p1_-| tbb3)  -- ([xshift=0mm]tbb3.south);

%% robot1
	\path[-latex, thick]
		 	(r0ap2) 		edge [pre]                            (r0p1)
			([yshift=-2mm]r0ap2.north west) edge [pre]                  ([yshift=-2mm,xshift=1mm]r0t0.north east)
					edge [post]                 ([yshift=-2mm,xshift=1mm]r0t0.north east)
			(r0ap2)		edge [post]                            (r0d0p1_)
					edge [post, bend right=10]                            (check00)
		 	(r0bp2)		edge [pre, bend right=10]                            (r0p1)
					edge [pre]                            (r0t0)
					edge [post]                            (r0t1.30)
					edge [post]                            (r0d1p1_)
			(r0bp2.east)edge [post]                            (check01)
		 	(r0cp2)		edge [pre]                            (r0p1)
			([yshift=2mm]r0cp2.south west)		edge [pre]                     ([yshift=2mm,xshift=1mm]r0t1.south east)
					edge [post]                            ([yshift=2mm,xshift=1mm]r0t1.south east)
			(r0cp2)		edge [post, bend left=0]                            (r0d1p1_)
					edge [post]                            (check01)
		 	(r0dp2) 	edge [pre,bend right=20]                            (r0p1)
					edge [pre]                            (r0t1)
					edge [post]                            (r0t0.-30)
			(r0dp2.north east)	edge [post, bend left=0]                            (r0d0p1_)
			(r0dp2)		edge [post, bend right=30]                            (check00)
			(badC01)	edge [pre]				(check01)
					edge [post]				(bad4);
%% robot2
	\path[-latex, thick]
		 	(r1ap2) 		edge [pre]                            (r1p1)
			([yshift=-2mm]r1ap2.north west) edge [pre]                  ([yshift=-2mm,xshift=1mm]r1t0.north east)
					edge [post]                 ([yshift=-2mm,xshift=1mm]r1t0.north east)
			(r1ap2)		edge [post]                            (r1d0p1_)
					edge [post,bend right=20]                            (check10)
		 	(r1bp2)		edge [pre]                            (r1p1)
					edge [pre]                            (r1t0)
					edge [post]                            (r1t1)
					edge [post]                            (r1d1p1_)
			(r1bp2.-20)		edge [post, bend left=10]                            (check11)
		 	(r1cp2) 	edge [pre]                            (r1p1)
			([yshift=2mm]r1cp2.south west)		edge [pre]                            ([yshift=2mm,xshift=1mm]r1t1.south east)
					edge [post]                            ([yshift=2mm,xshift=1mm]r1t1.south east)
			(r1cp2)		edge [post, bend left=5]                            (r1d1p1_.-75)
					edge [post, bend left=0]                            (check11)
		 	(r1dp2)		edge [pre, bend right=10]                            (r1p1)
					edge [pre]                            (r1t1)
					edge [post]                            (r1t0)
			(r1dp2.20)	edge [post, bend right=5]                            (r1d0p1_.-110)
			(r1dp2.west)		edge [post,bend right=20]                            (check10)
			(badC11)	edge [pre]				(check11)
					edge [post]				(bad5);

%%%%%%%%%%%%%%% check 
	\path[-latex, thick]
		 	(c1) 		edge [pre]                            (tools1)
					edge [pre]				(check10)
			(c1.south)	edge [post]				(res2)
		 	(c2) 		edge [pre]                            (tools1)
					edge [pre]				(check00)
					edge [post]				(res1)
;
	\path[-latex, thick]
		 	(c3) 		edge [pre]                            (tools2)
					edge [pre]				(check11)
					edge [post]				(res2)
		 	(c4) 		edge [pre]                            (tools2)
					edge [pre]				(check01)
			(c4.south)		edge [post]				(res1)
;
	\path[-latex, thick]
		 	(nxt) 		edge [pre]                            (c)
					edge [pre]                            (res1)
					edge [pre]                            (res2)
					edge [post]				(tools1)
					edge [post]				(tools2)
					edge [post, bend right=50]		(work);

	\draw [-{latex}, thick] (nxt.east)  -- ++(96mm,0)  -- ++(0,255mm) -- ++(-99mm,0) --        (env);
\end{tikzpicture}